\documentstyle[psfig]{mn2e}
\begin{document}

\title[Stochastic accretion of planetesimals onto white dwarfs]
{Stochastic accretion of planetesimals onto white dwarfs:
constraints on the mass distribution of accreted material from
atmospheric pollution}
\author[M. C. Wyatt et al.]
  {M. C. Wyatt$^1$\thanks{Email: wyatt@ast.cam.ac.uk},
   J. Farihi$^1$\thanks{STFC Ernest Rutherford Fellow},
   J. E. Pringle$^1$,
   A. Bonsor$^{2,3}$,
\\
  $^1$ Institute of Astronomy, University of Cambridge, Madingley Road,
  Cambridge CB3 0HA, UK\\
  $^2$ Institut de Plan\'{e}tologie et d'Astrophysique de Grenoble,
       Universit\'{e} Joseph Fourier, CNRS, BP 53, 38041 Grenoble, France\\
  $^3$ H.H. Wills Physics Laboratory, University of Bristol,
       Tyndall Avenue, Bristol BS8 1TL, UK
}

\maketitle

\begin{abstract}
This paper explores how the stochastic accretion of planetesimals onto white dwarfs would
be manifested in observations of their atmospheric pollution.
Archival observations of pollution levels for unbiased samples of DA and non-DA white dwarfs
are used to derive the distribution of inferred accretion rates, confirming that rates become
systematically lower as sinking time
(assumed here to be dominated by gravitational settling)
is decreased, with no discernable dependence on cooling age.
The accretion rates expected from planetesimals that are all the same mass (i.e., a
mono-mass distribution)
are explored both analytically and using a Monte Carlo model, quantifying how measured
accretion rates inevitably depend on sinking time, since different sinking times probe
different times since the last accretion event.
However, that dependence is so dramatic that a mono-mass
distribution can be excluded within the context of this model.
Consideration of accretion from a broad distribution of planetesimal
masses uncovers an important conceptual difference:
accretion is continuous (rather than stochastic) for planetesimals below a certain
mass, and the accretion of such planetesimals determines the rate typically inferred from
observations;
smaller planetesimals dominate the rates for shorter sinking times.
A reasonable fit to the observationally inferred accretion rate distributions is
found with model parameters consistent with
a collisionally evolved mass distribution up to Pluto-mass, and an underlying
accretion rate distribution consistent with that expected from descendants of
debris discs of main sequence A stars.
With these parameters, while both DA and non-DA white dwarfs accrete from the
same broad planetesimal distribution, this model predicts that the
pollution seen in DAs is dominated by the continuous accretion of $<35$~km
objects, and that in non-DAs
by $>35$~km objects (though the dominant size varies between stars by around an order of
magnitude from this reference value).
Further observations that characterise the dependence of inferred accretion rates
on sinking time and cooling age (including a consideration of the effect of
thermohaline convection on models used to derive those rates), and the decadal
variability of DA accretion signatures, will improve constraints on the mass distribution
of accreted material and the lifetime of the disc through which it is accreted.
\end{abstract}


\section{Introduction}
\label{s:intro}
Our understanding of the planetary systems around main sequence Sun-like stars has
grown enormously in the past few years.
Not only do we know about planets like Jupiter orbiting $0.05-5$~AU from their
stars, but a new population of low mass planets ($2-20$ times the mass of Earth)
orbiting within 1~AU has been found in transit and radial velocity surveys, as well
a more distant $8-200$~AU population of giant planets found in imaging surveys
(Udry \& Santos 2007).
Our understanding of the debris discs, i.e. belts of planetesimals and dust,
orbiting main sequence stars has also grown rapidly;
surveys show that $>50$\% of early-type stars host debris (Wyatt 2008).
Most of this debris lies $\gg 10$~AU in regions analogous to the
Solar System's Kuiper belt, but a few \% of stars exhibit dust at $\sim 1$~AU
that may originate in an asteroid belt analogue.

Much less is known about the planetary systems and debris of post-main sequence
stars, though these should be direct descendants of the main sequence population.
Several post-main sequence planetary systems are now known (e.g., Johnson et al. 2011),
but the debris discs of post-main sequence stars have remained elusive (though there
are examples around subgiants, e.g., Bonsor et al. 2013).
The closest to a counterpart of the Kuiper belt-like discs found around main
sequence stars may be the $30-150$~AU disc at the centre of the Helix nebula
(Su et al. 2007) and a few others like it (Chu et al. 2011; Bilikova et al. 2012).
However, a more ubiquitous phenomenon is that a large fraction of cool ($<25,000$K) 
white dwarfs show metals in their atmospheres.
This is surprising because their high surface gravities and small (or non-existent)
convection zones mean that such metals sink on short (day to Myr) timescales
implying that material is continuously accreted onto the stars with \textit{polluted}
atmospheres.
It has been shown that this material does not originate from the interstellar
medium (Farihi et al. 2009; 2010), and its composition
has been derived from atmospheric abundance patterns to be similar to
terrestrial material in the Solar System (Zuckerman et al. 2007; Klein et al. 2010;
G\"{a}nsicke et al. 2012).
The prevailing interpretation is that asteroidal or cometary material
is being accreted from a circumstellar reservoir, i.e., from the remnants of the
star's debris disc and/or planetary system.

Meanwhile a complementary set of observations provides clues to the accretion
process, since around 30 white dwarfs also show near-IR emission from dust
(Zuckerman \& Becklin 1987; Graham et al. 1990; Reach et al. 2005) and sometimes
optical emission lines of metallic gas (G\"{a}nsicke et al. 2006; Farihi et al. 2012a;
Melis et al. 2012) that is located within $\sim 1R_\odot$ from the stars.
Given its close proximity to the tidal disruption radius, and the fact that all white
dwarfs with evidence for hot dust or gas also show evidence for accretion in their
atmospheric composition, it is thought that both the dust, gas and atmospheric
pollution all arise from tidally disrupted planetesimals (Jura 2003).
However, the exact nature of the disc formation process, and
of the accretion mechanism are debated, which could for example be through
viscous processes or radiation forces (e.g., Rafikov 2011;
Metzger, Rafikov \& Bochkarev 2012).
It is also debated whether the pollution is caused by a continuous rain of small
rocks (Jura 2008), or by the stochastic accretion of much larger objects
(Farihi et al. 2012b).

In this paper we present a simple model of the accretion of
planetesimals in multiple accretion events to explore how such events are
manifested in observations of the star's atmospheric metal abundance.
The aim is to understand how such observations can be used to derive information
about the mass (or mass distribution) of accreted objects, and about whether
metal-polluted atmospheres are the product of steady state accretion of multiple objects
or the accretion of single objects.
A central motivation for this study is the recent claim that the distribution of
inferred accretion rates is different toward stars with different principal
atmospheric compositions (Girven et al. 2012; Farihi et al. 2012b),
and we show how this is an important clue
to determining the accretion process.
While others have recently shown that the previously unmodelled stellar process
of thermohaline convection can lead to substantial revision in the accretion rates
inferred toward some white dwarfs, potentially removing the difference in the 
inferred accretion rate distributions between the two populations (Deal et al. 2013),
we show here that such a difference is not unrealistic, rather it is almost unavoidable
within the context of the model presented here.

In \S \ref{s:obs} we compile observations from the literature and use these
to derive the distribution of inferred accretion
rates\footnote{Note that the rates we use here do not include the
effect of thermohaline convection, the effects of which have yet to be
fully characterised in this context.}
toward white dwarfs of different atmospheric properties (notably with different
sinking times for metals to be removed from the atmosphere) and ages.
A simple model is then presented in \S \ref{s:mod} that quantifies what we would
expect to observe if the planetesimals being accreted onto the white dwarfs all have the
same mass;
\S \ref{s:mono} demonstrates that such a model is a poor fit to the observationally inferred
accretion rate distributions, even if different stars are allowed to have different accretion rates and if the
model is allowed to include a disc lifetime that moderates the way accretion
is recorded on stars with short sinking times.
In \S \ref{s:mod2} the model is updated to allow stars to accrete material with
a range of masses, showing that this provides a much better 
fit to the observationally inferred accretion rate distributions.
The results are discussed in \S \ref{s:disc} and conclusions given in \S \ref{s:conc}.

\section{Distribution of accretion rates inferred from observations}
\label{s:obs}
The accretion rate onto a white dwarf can be inferred from observations
of its atmosphere, since its thin (or non-existent) convection zone means that
a metal (of index $i$) sinks on a relatively short timescale $t_{\rm{sink}(i)}$.
The exact sinking timescale depends on the metal in question and the properties
of the star, but can be readily calculated (e.g., Paquette et al. 1986).
In this paper the sinking process is assumed to be gravitational settling,
and so the sinking timescale is the gravitational settling timescale.
However, to allow for the possibility that other processes act to remove
metals from the convective zone (such as thermohaline convection), or indeed
to replenish it (e.g., radiative levitation), we refer to sinking timescales
rather than gravitational settling timescales throughout.

Thus observations of photospheric absorption lines, which can be used to infer 
the abundance of an element at the stellar surface and by inference 
the total mass of that element in the convection zone $M_{\rm{cv}(i)}$,
can be converted into an \textit{inferred mass accretion rate}
(assuming steady state accretion, Dupuis et al. 1992, 1993a, 1993b) of
\begin{equation}
  \dot{M}_{\rm{obs}(i)}=M_{\rm{cv}(i)}/t_{\rm{sink}(i)}.
  \label{eq:mdotobsi}
\end{equation}
Note that $\dot{M}_{\rm{obs}(i)}$ is expected to differ significantly
from the actual accretion rate, depending on the time variability of the accretion,
as outlined in this paper;
thus we use $\dot{M}_{\rm{obs}(i)}$ primarily as a
more convenient way of expressing $M_{\rm{cv}(i)}/t_{\rm{sink}(i)}$.
Measurements of different elements provide information on the composition of
the accreted material, which generally looks Earth-like (Zuckerman et al. 2007;
Klein et al. 2010; G\"{a}nsicke et al. 2012),
and extrapolation to any undetected metals can be used
to infer a total accretion rate $\dot{M}_{\rm{obs}}$.
It is worth emphasising that these accretion rates are not direct observables,
rather they need to be derived from stellar models (to get both $M_{\rm{cv}(i)}$
and $t_{\rm{sink}(i)}$).
As such, changes in stellar models can potentially lead to significant changes in
inferred accretion rates (e.g., Deal et al. 2013).
The models we use in \S \ref{ss:tsink} are those most commonly employed in the
white dwarf literature, though these have yet to incorporate the effects
of thermohaline convection.

Although the literature includes many studies that measure accretion rates towards
white dwarfs (e.g., Fig. 8 of Girven et al. 2012), for our purposes we will require
the distribution of accretion rates, i.e., the fraction of white dwarfs that exhibit
accretion rates larger than a given value $f(>\dot{M}_{\rm{obs}})$, for which
information about non-detections is as important as that about detections.
Thus here we perform a uniform analysis of data available in the literature
for samples chosen to be unbiased with respect to the processes that
may be causing atmospheric pollution.

From the outset it is important to note that this paper will distinguish between
two different atmospheric types:
DA white dwarfs that have H-dominated atmospheres, and non-DA white dwarfs
(comprised of basic sub-types DB and DC) that have He-dominated atmospheres.
This distinction is necessary, because metals have very different sinking times in
the two different atmospheres, and observations toward co-eval DA and non-DA white dwarfs
have different sensitivities to convection zone mass.
This distinction is discussed further in \S \ref{ss:dadb},
then \S \ref{ss:tsink} describes the uniform analysis employed, \S \ref{ss:samples}
describes the unbiased DA and non-DA samples, and the distributions
of accretion rates inferred from the observations are described in \S \ref{ss:distrn},
while \S \ref{ss:thermohaline} discusses uncertainties in the inferred accretion
rate distributions from the choice of model used to derive those rates.

\subsection{DA vs non-DA stars}
\label{ss:dadb}
An implicit assumption adopted here is that populations of both DA and
non-DA white dwarfs undergo the same history of mass input rate into the convection
zone;
i.e., two white dwarfs that are the same age can have different mass input rates, but
the distribution of mass input rates experienced by white dwarfs of the same age is
independent of their atmospheric type.
There are several channels by which both DA and non-DA white dwarfs might form.
However, most white dwarfs with He-dominated atmospheres (i.e., the non-DAs) are thought
to form from very efficient H-shell burning in the latter stages of post-main sequence
evolution, or late thermal pulses that dilute the residual H-rich envelope with metal-rich
material from the interior (e.g., Althaus et al. 2010).
So, as long as these processes are not biased in terms of stellar mass, or in terms
of planetary system properties, then it is reasonable to expect that the parent stars
(and circumstellar environments) of DA and non-DA white dwarf populations should be similar.
Indeed, observationally the mean mass of DB white dwarfs is very close to that
of their DA counterparts (e.g., Bergeron et al. 2011), though a small difference
has recently been discerned with DBs being slightly more
massive ($0.65M_\odot$ versus $0.60M_\odot$; Kleinman et al. 2013).
The low ratio of DB to DA white dwarfs in globular clusters (Davis et al.
2009) also suggests that the two populations could have different 
distributions of formation environments;
our assumption requires that this difference does not significantly
affect the planetary system properties (Zuckerman et al. 2010).
Practically, this assumption means that we expect the observationally inferred distribution
of accretion rates, $f(>\dot{M}_{\rm{obs}})$, to depend both on stellar age (because
of evolution of the circumstellar material) and on sinking time (because that affects
how the accretion rate is sampled), but not on the details of whether the star is a DA
or a non-DA.

\subsection{Uniform analysis}
\label{ss:tsink}
The uniform analysis consists of using reported measurements of atmospheric
Ca/H (for DAs) or Ca/He (for non-DAs) for stars for which their effective
temperature $T_{\rm{eff}}$ is also known.
These abundance measurements had been derived from modelling of stellar spectra
and were multiplied by the total convection zone mass (or that in the envelope above
an optical depth $\tau_R=5$; Koester 2009) to get the mass of Ca in that region.
The effective temperature is used to determine the sinking
timescale of Ca due to gravitational settling, $t_{\rm{sink(Ca)}}$,
for the appropriate atmospheric type using the
models of Koester (2009), and then the convection zone mass is converted into a mass
accretion rate of Ca.
This rate is scaled up by assuming that the Ca represents $1/62.5$ of the total
mass of metals accreted, like the bulk Earth, which appears broadly supported by data for
stars with Ca, Fe, Mg, Si, O and other metals detected (Zuckerman et al. 2010).

\begin{figure}
  \begin{center}
    \vspace{-0.1in}
    \begin{tabular}{c}
      \psfig{figure=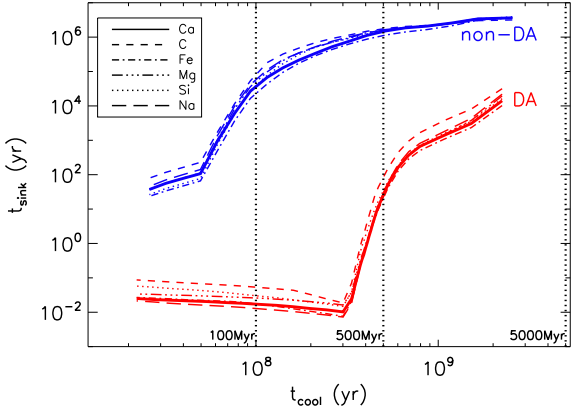,height=2.4in}
    \end{tabular}
    \caption{Sinking timescales due to gravitational settling at the base of the
    convection zone (or at an optical
    depth of $\tau_R=5$ if this is deeper) of different metals (shown with different
    line-styles
    as indicated in the legend) as a function of the star's cooling age (from tables 4-6 of
    Koester 2009) both for DA white dwarfs
    (i.e., those with H-dominated atmospheres, shown in red) and for non-DA white
    dwarfs (i.e., those with He-dominated atmospheres, shown in blue).
    We adopt the parameters for more efficient mixing in DAs cooler than 13,000K.
    }
   \label{fig:tsink}
  \end{center}
\end{figure}

The other parameter of interest is the star's cooling age $t_{\rm{cool}}$.
Although cooling age is actually a function of
$T_{\rm{eff}}$ and $\log{g}$, in practise the surface gravity is poorly known
due to insufficient observational data and a lack of good parallax measurements.
Thus throughout this paper we have assumed all stars
to be of typical white dwarf mass\footnote{Given the narrow distribution in
white dwarf masses estimated from gravitational redshifts (Falcon et al. 2010),
the uncertainty in cooling age from this assumption would be expected to be $<6$\%.}
with $\log{g}=8.0$, so that $T_{\rm{eff}}$ maps
uniquely onto a corresponding $t_{\rm{cool}}$, which also then maps onto a
corresponding $t_{\rm{sink}(i)}$.
Using this assumption, Fig.~\ref{fig:tsink} reproduces the sinking times due to
gravitational settling of a few metals
as a function of cooling age from Koester (2009) for both DAs and non-DAs.

Fig.~\ref{fig:tsink} shows that sinking times vary only by a factor of a few for
different metals in the same star, but that there is a large difference in
sinking timescale of a given metal when put in the atmosphere of the same star
at different ages, and for stars of the same age but of different atmospheric type.
For the DA white dwarfs $t_{\rm{sink}}$ can be as short as a few days
(e.g., Koester \& Wilken 2006), whereas for the non-DA white dwarfs
$t_{\rm{sink}}$ is more typically $0.01-1$~Myr (e.g. Koester 2009).
The dependence of sinking time on cooling age is similar for both atmospheric types
in that it is shorter at younger ages (i.e., at high effective temperatures),
followed by a transition to longer sinking times once temperatures are cool enough
for a significant convection zone to develop.

\subsection{DA and non-DA samples}
\label{ss:samples}
We consider two samples, one of DAs and the other of non-DAs.
The DA sample is comprised of 534 DA white dwarfs of which 38 have detections of
Ca, while the remaining 496 have upper limits on the presence of Ca.
These data comprise two surveys:
a Keck survey that specifically searched about 100 cool DA white dwarfs for Ca
absorption (Zuckerman et al. 2003), and 
the SPY survey which took VLT UVES spectra of $>500$ nearby white dwarfs
to search for radial velocity variations from double white dwarfs (SN Ia progenitors);
these data are also sensitive to atmospheric Ca (Koester et al. 2005).
The more accurate data were chosen in the case of duplication.
These stars are randomly chosen based on being nearby and bright,
and not biased in terms of the presence or absence of metals.

The non-DA sample is a small, but uniformly-sampled, set of DB stars searched
for metal lines with Keck HIRES (see Table 1 of Zuckerman et al. 2010).
Stars in this sample are predominantly young, with $50-500$~Myr cooling ages,
but are otherwise unbiased with respect to the likelihood to detect metal lines.
Although additional accretion rate measurements exist in the literature for DB stars,
these would only be suitable for inclusion in this study if the sample was unbiased
with regard to the presence of a disc, and if non-detections were reported with upper
limits on the accretion rates.

\begin{figure*}
  \begin{center}
    \vspace{-0.1in}
    \begin{tabular}{cccc}
      \hspace{-0.1in} \textbf{(a)} & \hspace{-0.3in}
      \psfig{figure=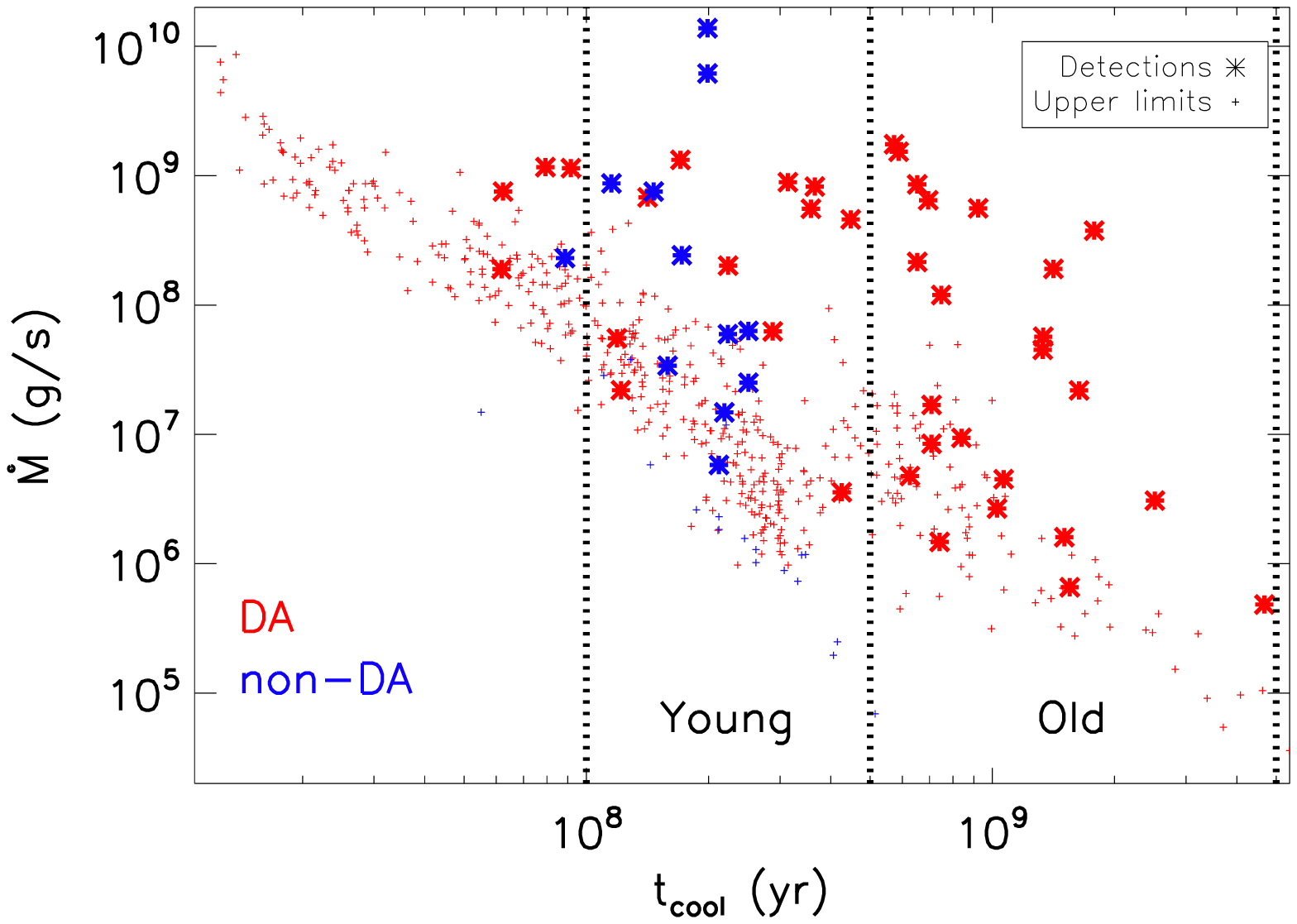,height=2.4in} &
      \hspace{-0.1in} \textbf{(b)} & \hspace{-0.3in}
      \psfig{figure=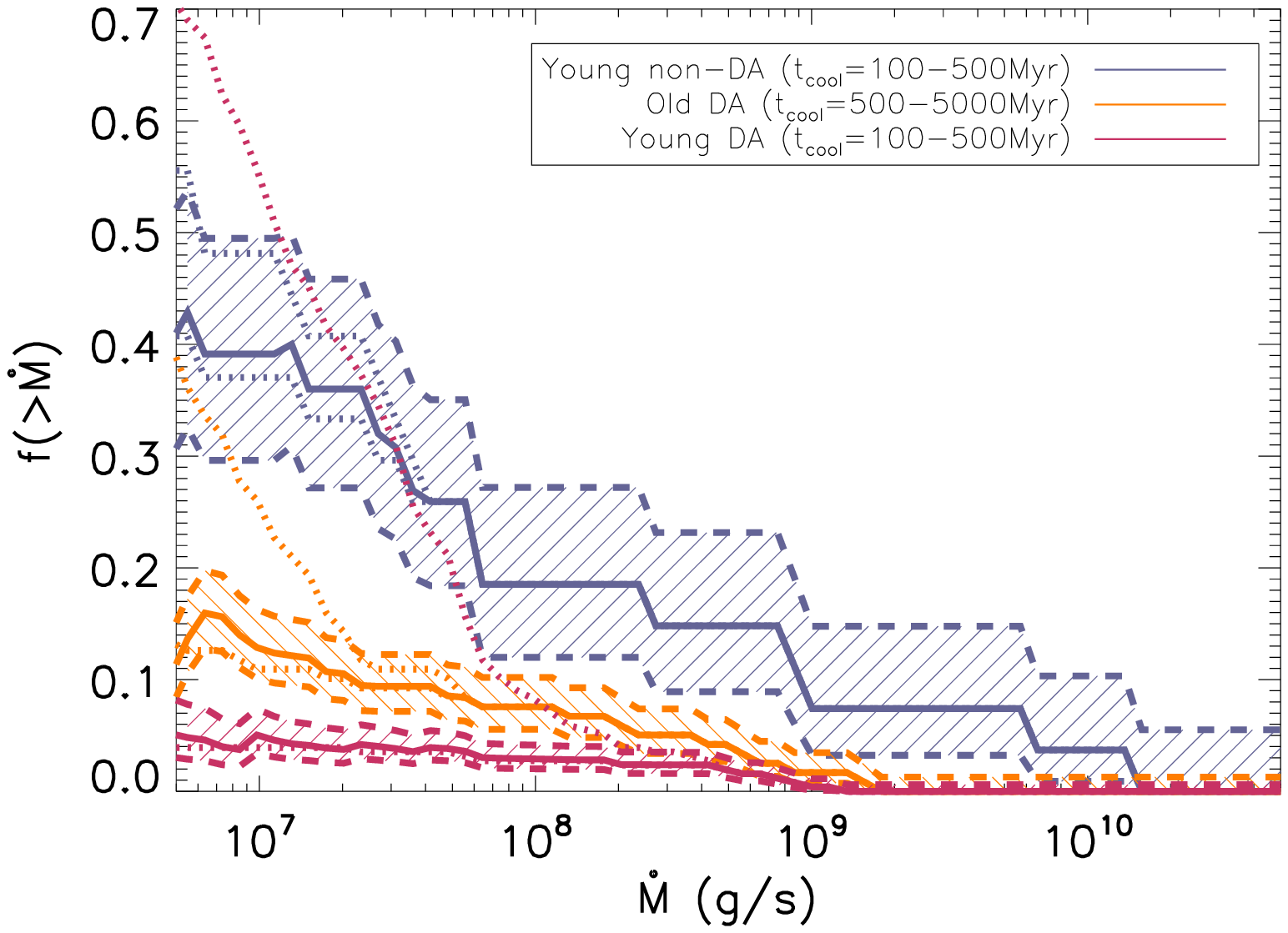,height=2.4in} \\
      \hspace{-0.1in} \textbf{(c)} & \hspace{-0.3in}
      \psfig{figure=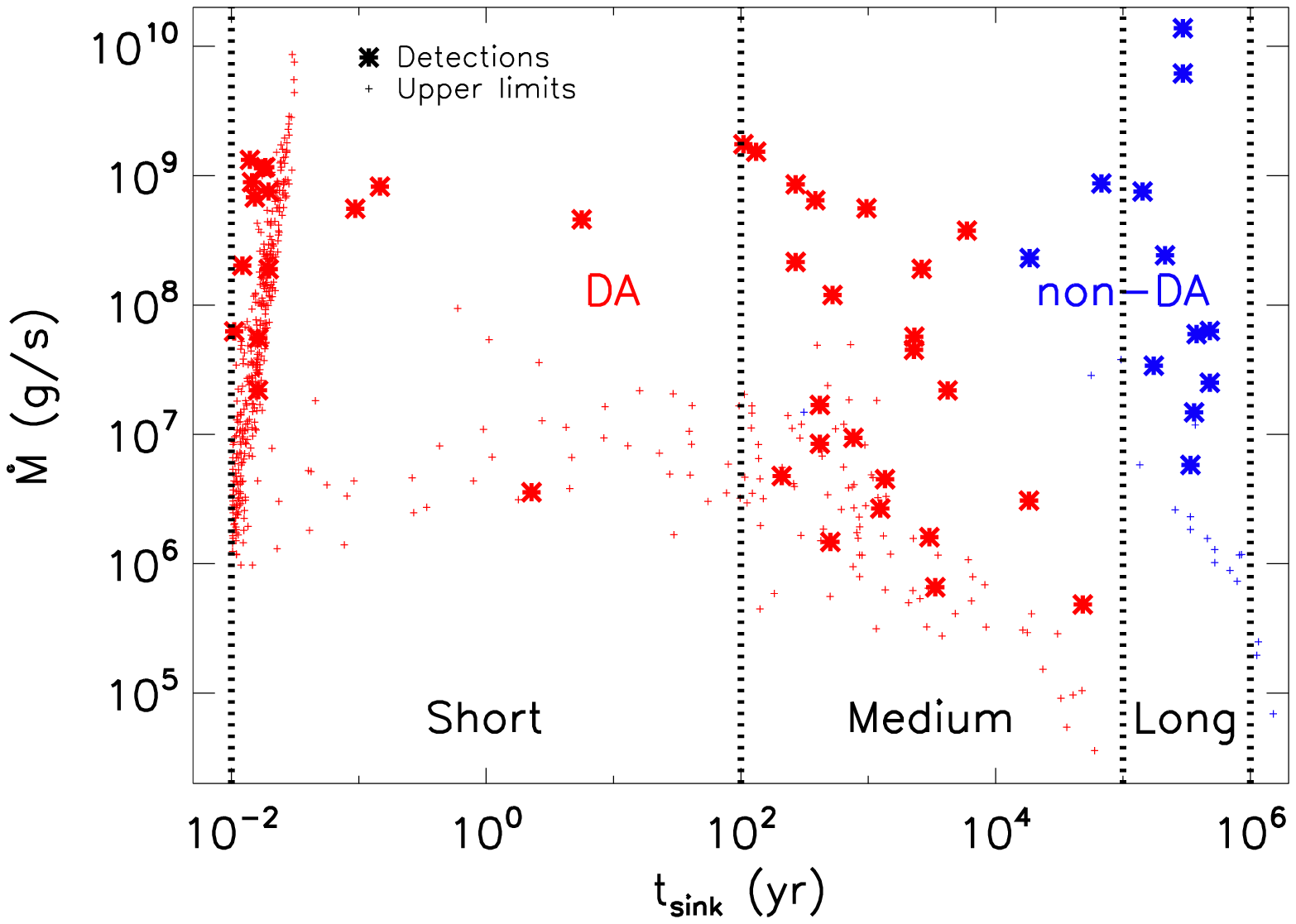,height=2.4in} &
      \hspace{-0.1in} \textbf{(d)} & \hspace{-0.3in}
      \psfig{figure=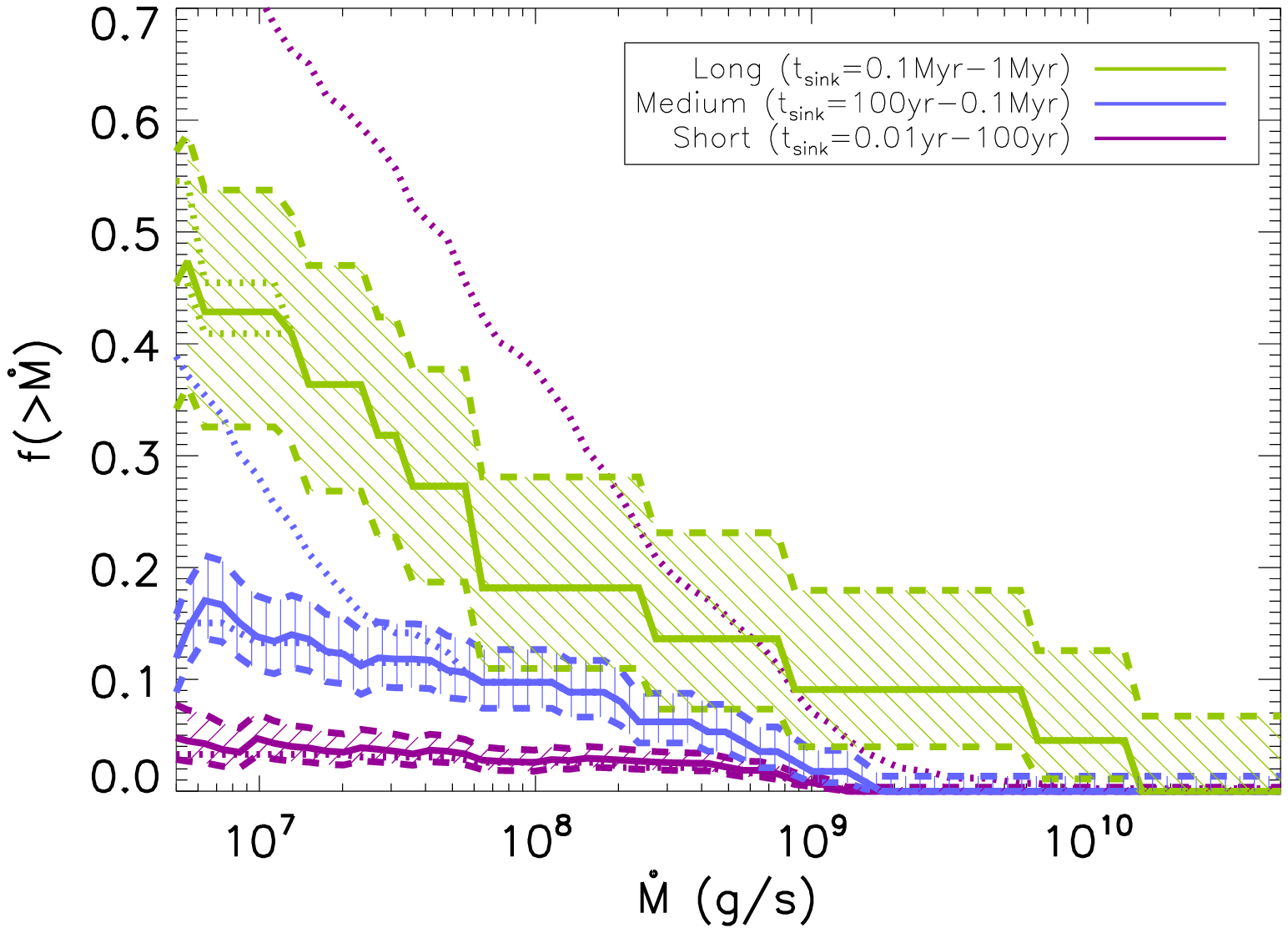,height=2.4in} \\
    \end{tabular}
    \caption{Inferred accretion rates for unbiased samples of DA white dwarfs
    (shown in red) and for non-DA white dwarfs (shown in blue).
    The left panels (\textbf{a} and \textbf{c}) show accretion rates inferred from
    Ca measurements assuming a terrestrial composition.
    Detections are shown with asterisks and upper limits with a small plus.
    In \textbf{(a)} the x-axis is the cooling age of the white dwarf inferred from the
    star's effective temperature (assuming $\log{g}=8.0$), whereas in \textbf{(c)}
    the x-axis is the sinking time of Ca inferred from the effective temperature.
    The right panels (\textbf{b} and \textbf{d}) show the fraction of white dwarfs in
    different sub-samples that have inferred accretion rates above a given level.
    These sub-samples are split by cooling age in \textbf{(b)} into young and
    old age bins, and by sinking time in \textbf{(d)} into short, medium and long sinking
    time bins; the bin boundaries are noted in the legends and 
    no distinction is made for the sub-samples in \textbf{(d)} between DAs and non-DAs.
    The dotted lines give the range of distributions inferred for each sub-sample for
    optimistic and pessimistic assumptions about the stars with upper limits (see text
    for details).
    The solid lines give the best estimate of the distributions for each sub-sample, and
    the dashed lines and hatched regions show the $1\sigma$ uncertainty due to small
    number statistics (see text for details).}
   \label{fig:obs}
  \end{center}
\end{figure*}

\subsection{Distribution of inferred accretion rates}
\label{ss:distrn}
The left panels of Fig.~\ref{fig:obs} show the inferred accretion rate data for the
two samples, plotted both against age (Fig.~\ref{fig:obs}a) and against sinking time
(Fig.~\ref{fig:obs}c).
The sense of the detection bias is evident from the lower envelope of the detections
in Fig.~\ref{fig:obs}a;
e.g., there are far fewer detections in the younger age bins due to the higher
temperature of these stars which makes Ca lines harder to detect for a given
sensitivity in equivalent width (see Fig. 1 of Koester \& Wilken 2006).

The right panels use the information in the left panels to determine the distribution
of inferred accretion rates $f(>\dot{M}_{\rm{obs}})$ for different sub-samples as
outlined in the captions.
For example, Fig.~\ref{fig:obs}b keeps the split between DA and non-DA and further sub-divides
these samples according to stellar age, using age bins of 100-500~Myr (here-on the young
bin) and 500-5000~Myr (here-on the old bin).
Fig.~\ref{fig:obs}d combines the DA and non-DA samples, but then makes sub-samples according
to sinking time bins of 0.01-100~yr (here-on the short bin), 100~yr-0.1~Myr (here-on the medium
bin) and 0.1-1~Myr (here-on the long bin), though
overlap between the DA and non-DA samples is confined to a small fraction (4.4\%) of non-DAs
in the medium bin.

Identifying the most accurate way to determine the underlying distribution of 
$f(>\dot{M}_{\rm{obs}})$ for the different sub-samples (i.e., that which would be measured with 
infinite sensitivity and sample size) is complicated by the fact that the observations only 
result in upper limits for many stars, and the sample size is finite,
a problem encountered many times in astrophysics though without a definitive solution
(e.g., Feigelson \& Nelson 1985; Mohanty et al. 2013).
Two bounds on the underlying distribution can be obtained by considering that the most pessimistic
assumption for the stars that have upper limits is that they are not accreting (i.e., that
with infinitely deep observations $\dot{M}_{\rm{obs}}=0$), while the most optimistic assumption
is that those stars are accreting at a level that is at the upper limit inferred from the
observations.
These bounds are plotted on Figs~\ref{fig:obs}b and \ref{fig:obs}d for the different sub-samples
with dotted lines, and one might expect the underlying distributions to fall between these two bounds.
However, while instructive, these bounds encounter two problems.
First, the optimistic limit requires the improbable occurrence of many detections at the
3$\sigma$ limit.
This problem is particularly acute when a significant fraction of the sample only has upper limits,
such as the short sinking time sub-sample on Fig.~\ref{fig:obs}d, because not only is it
statistically unlikely that the observer recorded an upper limit for each star when the true
accretion level was as high as assumed in the optimistic case, but also the small number of
actual detections already suggests that only a small fraction of stars should have detections
at such a high level.
In other words, the optimistic limit is unrealistically optimistic.
The second problem is that this does not account for small number statistics, which affects in 
particular the distribution at high accretion rates, where the optimistic and pessimistic lines 
converge, but where the rates have been estimated from very few detections.

Here we adopt an alternative method for estimating $f(>\dot{M}_{\rm{obs}})$ that circumvents
these two problems.
The idea is that if we want to know the fraction of stars in a sub-sample of size $N_{\rm{s}}$
that have accretion levels above say $\dot{M}_{\rm{obs}}=10^7$~g\,s$^{-1}$, then we should only
consider the subset of $N_{\rm{ss}}$ stars within that sub-sample for which accretion could
have been detected at that level.
The fraction of stars with accretion above that level is then the number of detections in that
subset $N_{\rm{ssdet}}$ (noting that this may be lower than the number of stars in the whole
sub-sample with accretion above that level) divided by $N_{\rm{ss}}$.
The uncertainty on that fraction can then be determined from $N_{\rm{ssdet}}$ and $N_{\rm{ss}}$
using binomial statistics (see Gehrels 1986), and it is evident that small number statistics
will be important both for large accretion rates where there are few detections
(small $N_{\rm{ssdet}}$), and for small accretion rates where few of the sub-sample can be
detected at such low levels (small $N_{\rm{ss}}$).
In Figs~\ref{fig:obs}b and \ref{fig:obs}d we show the fraction determined in this way
with a solid line, and the hatched region and dashed lines indicate the $1\sigma$ uncertainty.
\footnote{Note that these errors apply only to the measurement of $f(>\dot{M}_{\rm{obs}})$
at a specific accretion rate and so the points on this line are not independent of each other.
This is relevant when assigning a probability that a given model provides a good fit to the data,
as will be discussed later.}
This method only works as long as stars are included in the subset in a way that does not introduce
biases with respect to the level of accretion.
In this case Fig.~\ref{fig:obs}a shows that as we try to measure the distribution down to
lower levels of accretion, the only bias is that the subset becomes increasingly biased
toward the older stars in the sub-sample.
So, the distribution we infer in this way is only a good representation of that of the 
whole sub-sample as long as the inferred accretion rate distribution is not strongly dependent 
on cooling age, a topic we address below.

While Figs~\ref{fig:obs}b and \ref{fig:obs}d provide the best estimate of the underlying
inferred accretion rate distributions in the sub-samples, we will also use Fisher's exact test to
assign a probability to the null hypothesis that two sub-samples have the same inferred accretion
rate distribution.
To do so we just need four numbers, $N_{\rm{ssdet}}$ and $N_{\rm{ss}}$ for the
two sub-samples measured at an appropriate accretion level, and the probability quoted
will be that for the observations of these sub-samples resulting in rates that are
as extreme, or more extreme, if the null hypothesis were true.

The first thing to note from Fig.~\ref{fig:obs}b is that the distributions of inferred accretion
rates in the young age bin are significantly different between the DA and non-DA populations.
For example, for the subsets corresponding to accretion above $10^7$~g\,s$^{-1}$, there is
only a 0.002\% probability of obtaining rates as extreme as, or more extreme than, the
4.6\% (6/131) of young DAs and the 39\% (9/23) of young non-DAs if the two are drawn from the
same distribution.
If as assumed in \S \ref{ss:dadb} the only difference between the underlying distribution of inferred
accretion rates toward these stars is the sinking timescale on which the accretion rate is measured,
then this indicates that the longer sinking times of the non-DA population (with a median level of
0.37~Myr) have lead to a distribution with higher inferred accretion rates than the DA
population (with a median sinking time of 5~days).

Concentrating now on the inferred accretion rate distributions for the DA sub-samples in
Fig.~\ref{fig:obs}b we conclude that there is no strong evidence that these vary with age.
For example, taking again subsets corresponding to accretion above $10^7$~g\,s$^{-1}$, there is
a 2.6\% probability of obtaining rates as extreme as, or more extreme than, the
4.6\% (6/131) of young DAs and the 12.4\% (13/105) of old DAs if the two are drawn from the
same distribution.
While the small difference in rates between the populations could be indicative of an
age dependence in the inferred accretion rate (higher rates around older stars), this is of low
statistical significance.
Moreover, since age is correlated with sinking time in the DA sub-samples (Fig.~\ref{fig:tsink}),
and the previous paragraph concluded that longer sinking times lead to higher inferred
accretion rates, it is possible that the (marginally) higher accretion
rates around the older DA sub-sample are due to their longer sinking times relative
to the younger DA sub-sample, and have nothing to do with the evolution of the underlying
accretion rate distribution.
However, it is not possible to conclude that age is not an important factor in determining
the inferred accretion rate distribution, as there could even be a strong decrease in
accretion rate with age that has been counteracted in the sub-samples of Fig.~\ref{fig:obs}b
by the sinking time dependence.
To assess the effect of age properly would require comparison of sub-samples of DAs and non-DAs
with the same sinking times but different ages, but this is not available to us for now
(see Fig.~\ref{fig:obs}c).
Nevertheless, since we do not see any evidence for a dependence on age
(see also Koester 2011), our analysis in this paper
will assume the underlying distribution of accretion rates to be independent of age
(noting that an age dependence in the distribution of inferred accretion rates may arise
through the sinking time).

Given that sinking time is likely the dominant factor, the most important plot is
Fig.~\ref{fig:obs}d.
The picture that emerges reinforces the previous conclusion on the importance
of sinking time in the inferred accretion rate distributions, and furthermore points to a
monotonic change in the distribution of inferred accretion rates, with longer sinking times
resulting in higher inferred accretion rates.
To quantify the significance of the difference between the sub-samples, take again subsets
corresponding to accretion above $10^7$~g\,s$^{-1}$;
there is a 0.0006\% probability of obtaining levels as extreme as, or more extreme than,
the 43\% (9/21) rate in the long bin and the 4.3\% (6/140) rate in the short bin
if the two are drawn from the same distribution.
This probability becomes 0.4\% when comparing the rate in the long bin with the
13.4\% (13/97) rate in the medium sinking time bin,
and 1.1\% when comparing the rates in the short and medium sinking time bins
(this latter probability is further reduced to $\sim 0.6$\% if larger accretion rates up to
$10^{8}$~g\,s$^{-1}$ are considered).
That is, as expected from above, there is a significant difference between the sinking time
bins, though the confidence level that all three sinking time bins have distributions
that are different from each other, and hence that there is a monotonic change
in inferred accretion rates across a wide range of sinking times, is slightly below $3\sigma$.

While the above analysis is not sufficient to make a strong statement about the difference
between (say) the short and medium sinking time bins, we take the near $3\sigma$ significance
to indicate that future observations will soon be able to find such a difference, if it exists.
Thus we tailor the models in the following sections to reproduce as good a fit to the solid
lines in Fig.~\ref{fig:obs}d as possible.
This approach allows us demonstrate the qualitative behaviour of the models, and
how the different parameters affect their predictions for the dependence of the
inferred accretion rate distribution on sinking time.
However, in doing so we recognise that this approach may appear to constrain the model in ways that
will not be formally significant given the limitations of small number statistics, and
note in future sections where that is the case.

Note that while we have assumed that there is no dependence of accretion rate on age,
the lack of evolution is not well constrained, and the different sinking time bins have
different age distributions;
the median ages are 140, 840 and 220~Myr for the short, medium and long bins, respectively.
If there was a dependence of accretion rate on age, the most significant effect
would likely be on the position of the medium sinking time bin with respect to
the other bins.
For example, a decrease in accretion rates with age would mean the distribution
$f(>\dot{M}_{\rm{obs}})$ for the medium bin would be higher if plotted at a
comparable age to that of the long and short bins.

\subsection{Caveats}
\label{ss:thermohaline}
The method described above to derive accretion rates
makes some simplifications about the evolution of accreted metals.
Specifically the assumption is that metals are removed from the observable outer atmosphere over
a sinking time, where the sinking time is that due to gravitational settling.
This is the standard approach in the literature (e.g., Koester 2009).
However one important process that is omitted here is thermohaline (or fingering) convection.
Thermohaline convection is triggered by a gradient in metallicity in the stellar atmosphere that
decreases toward the centre, such as would be expected if high metallicity material had been
accreted at the surface.
In such a situation, the metals can be rapidly mixed into the interior through metallic
fingers, analogous to salt fingers studied in the context of Earth's oceans (e.g., Kunze 2003).
Application of this process to general astrophysical situations, such as mixing in
stellar atmospheres, has been characterised using 3D numerical simulations
(Traxler et al. 2011; Brown et al. 2013).
Thermohaline convection has been shown to have important consequences for
mixing of planetary material accreted by main sequence stars
(Vauclair 2004; Garaud 2011), for stars that accreted material from an AGB companion
(Stancliffe \& Glebeek 2008), and possibly for low mass RGB stars (Denissenkov 2010).

A recent study also found that this process may be important for 
accretion onto white dwarf atmospheres (Deal et al. 2013), in that accretion rates
inferred from observations of DA white dwarfs may actually be higher than previously
considered.
The rates for non-DA white dwarfs would be unaffected by this process leading to the interesting
possibility that the distribution of rates for both populations are the same.
However, for now the model has been only been applied to 6 stars, and the implications have yet
to be characterised across the range of stellar and pollution parameters required
in this study.
As such it is premature (and not possible with published information) to use
rates that account for thermohaline convection in this paper.
Nevertheless, since this process has the potential to affect inferred accretion rates,
and may also do so in a way that depends on sinking time, a caveat is required when
interpreting the conclusions in \S \ref{ss:distrn} about how accretion rate distributions
depend on sinking time.
If the rates need to be modified as a result of this process, the analysis in this paper
could be repeated, and we note below the potential implications if the rate was to turn out to
be independent of sinking time.

\section{Simple model: stochastic accretion of mono-mass planetesimals}
\label{s:mod}
The dependence of inferred accretion rates on sinking time has previously been noted by
Girven et al. (2012) and discussed further in Farihi et al. (2012b) from a difference
between the accretion rates inferred toward DA and non-DA populations.
It is interpreted as evidence of the stochastic nature of the accretion process,
with the short sinking time DAs providing a measure of the \textit{instantaneous}
level of accretion being experienced by the star, and the longer sinking time of non-DAs
providing evidence for \textit{historical} accretion events, such as the accretion of a
large comet which can leave mass in the atmospheres of non-DAs for long periods
after the event.
In this section we use a pedagogical model to illustrate the nature of stochastic
processes and to quantify how different mass accretion rates (of objects of finite
mass) would be expected to be inferred toward white dwarf populations with different
sinking times.

\subsection{Pedagogical model}
\label{ss:ped}
Consider a white dwarf at which planetesimals are being thrown at a
mean rate $\dot{M}_{\rm{in}}$.
Here it is assumed that all planetesimals have the same mass $m_{\rm{p}}$,
and that once accreted at time $t_i$, the mass from planetesimal $i$ that
remains potentially visible in observations of the white dwarf's atmosphere decays exponentially on the
sinking time $t_{\rm{sink}}$, i.e., for $t>t_i$
\begin{equation}
  \label{eq:matmi}
  m_{{\mathrm{atm}},i} = m_{\mathrm{p}} {\mathrm{e}}^{(t_i-t)/t_{\mathrm{sink}}}.
\end{equation}
Note that after being accreted the planetesimal is mixed nearly instantaneously within
the white dwarf's convective zone, and only a small fraction of that mass contributes
to the observable atmospheric signatures at any one time.
Thus by $m_{{\mathrm{atm}},i}$ we really mean the mass of planetesimal $i$ that
remains in the convective zone, which can be determined through observations
of abundances in the white dwarf's atmosphere using a stellar model to determine the
total mass of the convective zone over which that abundance is assumed to apply.

The total mass of pollutants that are present in the convective zone,
and hence potentially visible in the white dwarf's atmosphere at any one
time, which we call the atmospheric mass, is the sum of all previous
accretion events, depleted appropriately by the decay, i.e.,
\begin{equation}
  M_{\rm{atm}} = \sum_i m_{{\rm{atm}},i}.
  \label{eq:matm}
\end{equation}
We also define the accretion rate that would be inferred from such an
atmospheric mass as
\begin{equation}
  \dot{M}_{\rm{atm}} = M_{\rm{atm}}/t_{\rm{sink}}.
  \label{eq:mdotatm}
\end{equation}
Note the similarity with eq.~(\ref{eq:mdotobsi}), which is because we
will be comparing $\dot{M}_{\rm{atm}}$ with $\dot{M}_{\rm{obs}}$,
and underscores the importance of using the same value of $t_{\rm{sink}}$
in the modelling as that used to obtain accretion rates from the observations.

Since the mass can only arrive in units of $m_{\rm{p}}$, this is a Poisson
process, and $\dot{M}_{\rm{atm}}$ is not necessarily equal to $\dot{M}_{\rm{in}}$.
Rather the inferred accretion rate has a probability density
function $P(\dot{M}_{\rm{atm}})$, and an associated cumulative distribution
function that we characterise by
\begin{equation}
  f(>\dot{M}_{\rm{atm}})=\int_{\dot{M}_{\rm{atm}}}^\infty P(x) dx,
  \label{eq:fgtmdot}
\end{equation}
which is the fraction of the time we would expect to measure an accretion rate
larger than a given value.

The set-up of this problem is exactly the same as that for shot noise, the nature of
which depends on the parameter $n$, the mean number of shots per unit time
(see Appendix \ref{a:shot}).
For our problem, 
\begin{equation}
  n=\dot{M}_{\rm{in}}t_{\rm{sink}}/m_{\rm{p}}
  \label{eq:n}
\end{equation}
is the mean number of accretion events per sinking time, and the shots have
the form
\begin{equation}
  F(\tau) = H(\tau) \rm{e}^{-\tau},
  \label{eq:f}
\end{equation}
where $\tau=t/t_{\rm{sink}}$ is time measured in units of the sinking timescale,
$H(\tau)$ is the Heaviside step function, and the \textit{shot amplitude} discussed
in the appendix and references therein should be scaled by $m_{\rm{p}}/t_{\rm{sink}}$
to get this in terms of the inferred accretion rate.

Here we derive the cumulative distribution function using a Monte Carlo model (\S \ref{ss:monte}),
and apply results from the literature for shot noise to explain the shape of
the distribution function analytically (\S \ref{ss:analyt}).

\subsection{Monte Carlo model}
\label{ss:monte}
For a white dwarf with a given $t_{\rm{sink}}$, and accretion defined by
$\dot{M}_{\rm{in}}$ and $m_{\rm{p}}$, we first define a timestep
$dt=t_{\rm{sink}}/N_{\rm{sink}}$, where $N_{\rm{sink}}$ is the number of
timesteps per sinking time (this should be large enough to recover the shape
of the exponential decay of atmospheric mass, and is set to 10 here).
We then set a total number of timesteps, $N_{\rm{tot}}$ (set to 200,000 here),
and use Poisson statistics to assign randomly the number of planetesimals
accreted in each timestep (using the \textit{poidev} routine, Press et al. 1989, and a
mean of $\dot{M}_{\rm{in}}dt/m_{\rm{p}}$).
The $N_{\rm{tot}}$ timesteps are considered as a (looped) time series, and so
the mass accreted in each timestep is carried forward to subsequent timesteps
with the appropriate decay (eq.~\ref{eq:matmi}) to determine the mass in the
atmosphere and inferred accretion rate as a function of time.

\begin{figure}
  \begin{center}
    \vspace{-0.1in}
    \begin{tabular}{cc}
      \hspace{-0.2in} \textbf{(a)} & \hspace{-0.2in} 
      \psfig{figure=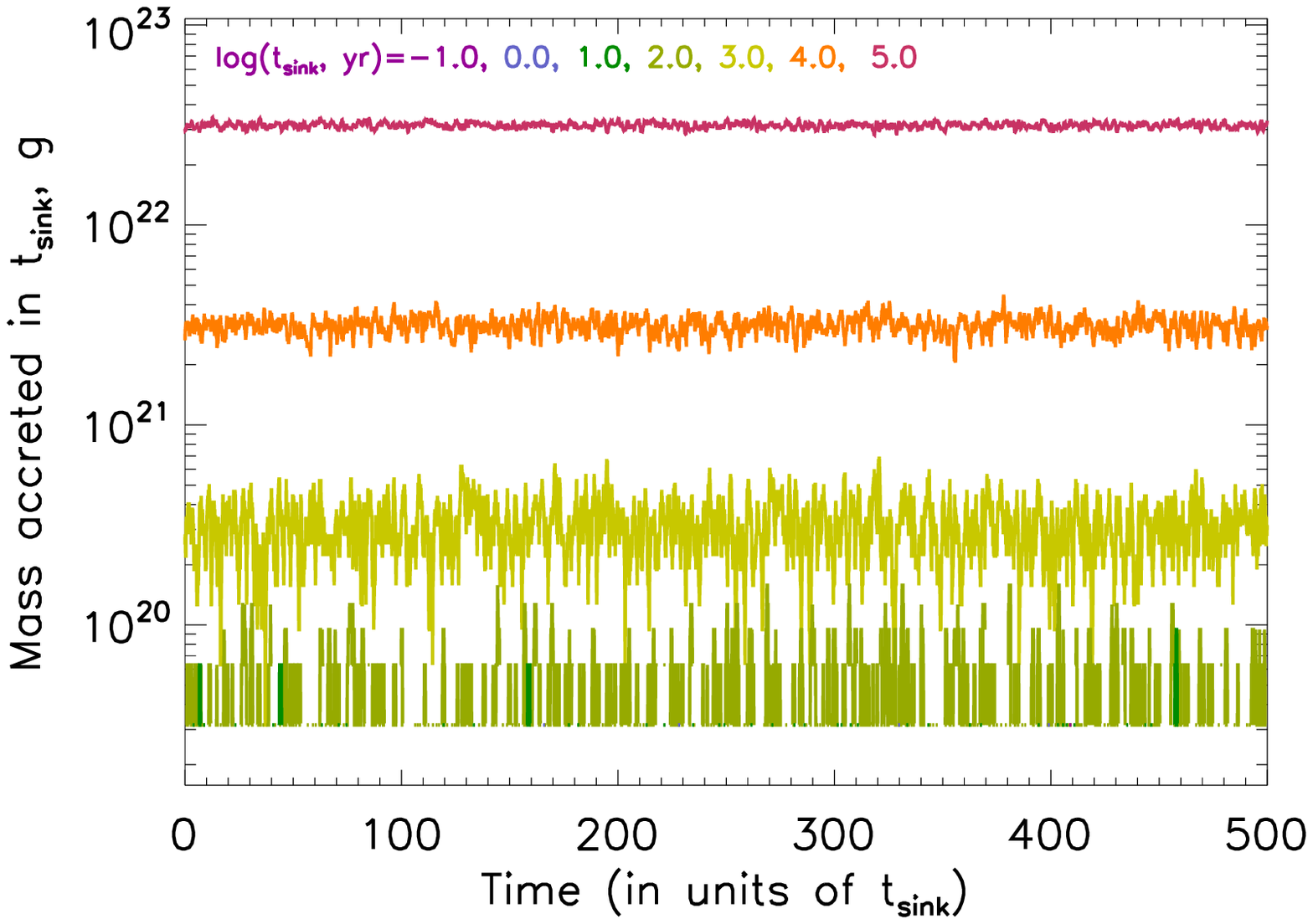,height=2.4in} \\
      \hspace{-0.2in} \textbf{(b)} & \hspace{-0.2in} 
      \psfig{figure=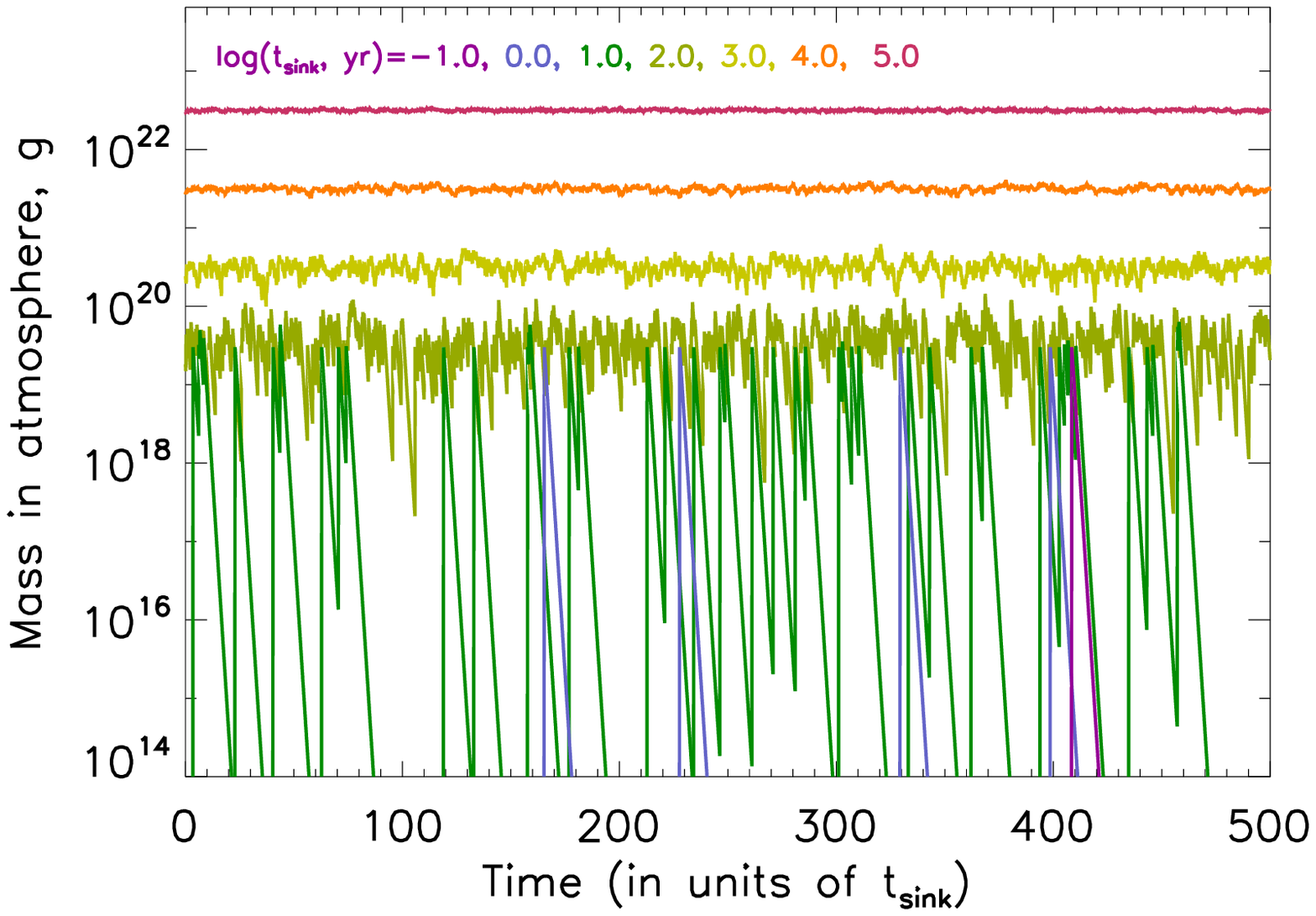,height=2.4in} \\
      \hspace{-0.2in} \textbf{(c)} & \hspace{-0.2in} 
      \psfig{figure=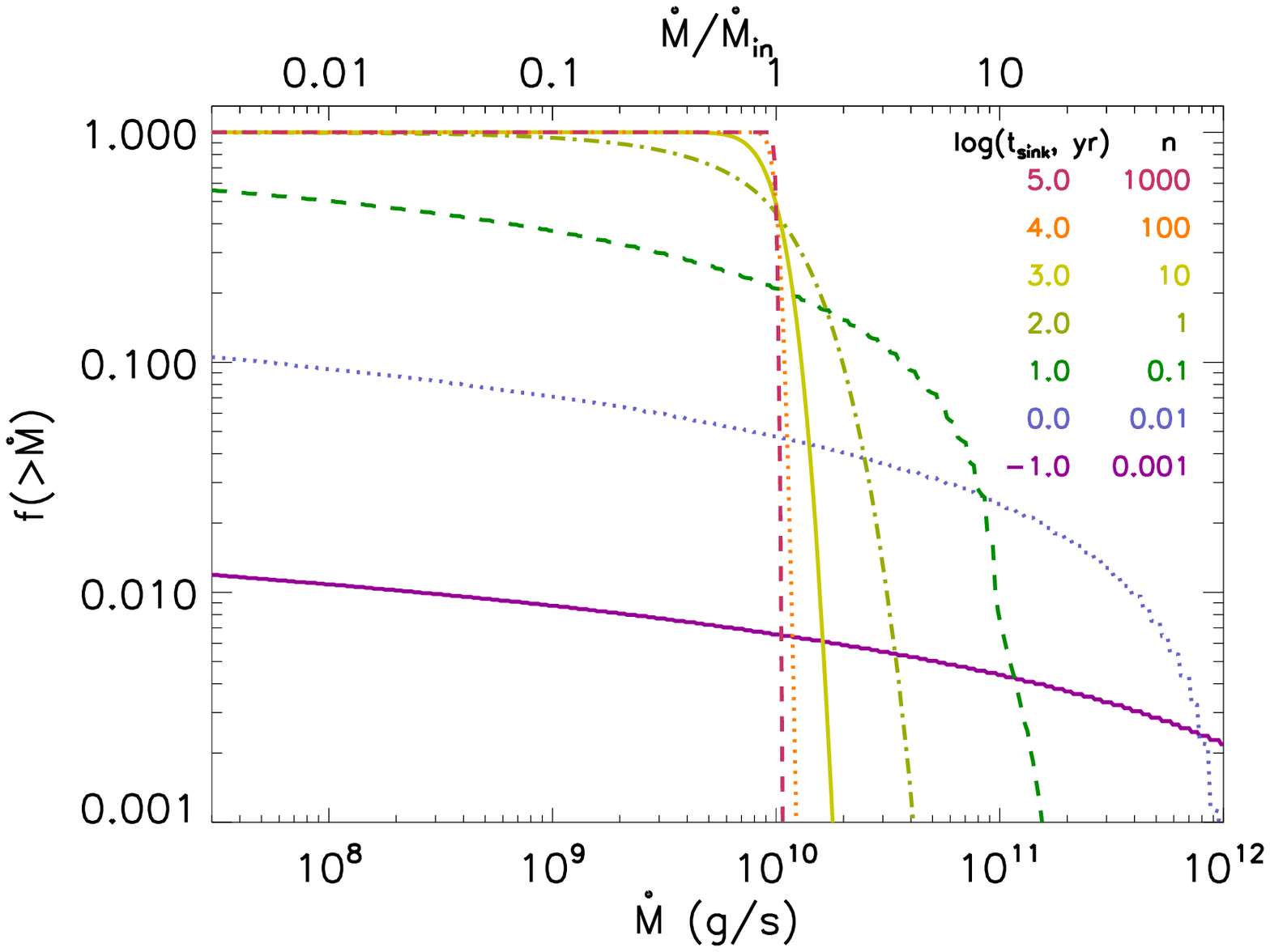,height=2.4in}
    \end{tabular}
    \caption{Monte carlo simulations of accretion of $3.2 \times 10^{19}$~g planetesimals
    at a mean rate $10^{10}$ g\,s$^{-1}$ onto white dwarfs with seven different
    sinking times $t_{\rm{sink}}$ logarithmically spaced between 0.1~yr and 0.1~Myr
    shown with different colours.
    \textbf{(a)} The total mass accreted in one sinking time, as a function of time,
    with only the first 500 sinking times shown for clarity.
    \textbf{(b)} The total mass remaining as potentially visible in the atmosphere as a
    function of time.
    \textbf{(c)} The fraction of all timesteps for which the accretion rate is measured
    to be above the rate given on the $x$-axis; i.e., the cumulative distribution
    function $f(>\dot{M}_{\rm{atm}})$.
    The top axis generalises this plot to dimensionless accretion
    rate ($\dot{M}/\dot{M}_{\rm{in}}$) when used in conjunction with the number of accretion
    events per sinking time ($n$) given in the legend.
    }
   \label{fig:mcmono}
  \end{center}
\end{figure}

Fig.~\ref{fig:mcmono} shows the result of this process for canonical parameters
of $\dot{M}_{\rm{in}}=10^{10}$ g\,s$^{-1}$ and $m_{\rm{p}}=3.2 \times 10^{19}$ g.
This accretion rate corresponds to the mass of the current asteroid belt
(Krasinsky et al. 2002) being accreted every $\sim 10$~Myr.
This planetesimal mass corresponds to a 27~km diameter planetesimal for a density
of 3~g\,cm$^{-3}$, and has been chosen so that a sinking time of 100~years
corresponds to a mean rate of one planetesimal being accreted per sinking time
(i.e., $n=1$).
This process has been repeated for seven different sinking times
that correspond to $n=0.001$, 0.01, 0.1, 1, 10, 100 and 1000
planetesimals being accreted per sinking time.

Fig.~\ref{fig:mcmono}a shows how longer sinking times (larger $n$) result in
larger quantities of mass accreted in one sinking time.
However, decreasing the sinking time runs into a barrier since the accreted
mass cannot be less than the mass of a single planetesimal.
Thus as $n$ is decreased to 1 and below, the mass accreted in any one sinking time
becomes more noticeably probabilistic.
The same effect is also seen in Fig.~\ref{fig:mcmono}b, except that the
mass remaining as potentially visible in the atmosphere can be less than $m_{\rm{p}}$.
Indeed for the shortest sinking times of 0.1 and 1 years, the atmospheric mass
spends most of its time at insignificantly small levels, increasing to the
level of $m_{\rm{p}}$ only immediately following an accretion event, with
exponential decay thereafter.
In constrast, for the longest sinking times ($n \gg 1$), the atmospheric mass
is approximately constant at a level $\dot{M}_{\rm{in}}t_{\rm{sink}}$.

The distribution of atmospheric masses is quantified in Fig.~\ref{fig:mcmono}c, which
shows the fraction of time the accretion rate would be inferred to be above a
given level.
For long sinking timescales ($n \gg 1$), this is close to a step function, transitioning
from 1 to 0 close to $\dot{M}_{\rm{in}}$; i.e., the inferred accretion rate is always
very close to the mean level.
For short sinking timescales ($n \ll 1$) however, the inferred accretion rate covers a
broad range, from around $m_{\rm{p}}/t_{\rm{sink}}$ just after an accretion event,
which is significantly higher than $\dot{M}_{\rm{in}}$ in this regime, down to levels far
below $\dot{M}_{\rm{in}}$.
As noted in \S \ref{ss:analyt}, the distribution at levels just below
$m_{\rm{p}}/t_{\rm{sink}}$ in this regime is to a reasonable approximation dictated
by the exponential decay function, since intermediate accretion rates are simply
the vestiges of earlier accretion events.

\subsection{Analytical}
\label{ss:analyt}
The distribution of shot noise characterised in the manner of equations (\ref{eq:n})
and (\ref{eq:f}) is given in section 6.1 of Gilbert \& Pollack (1960)
(see Appendix~\ref{a:shot}).
There they derive the exact form of the probability density distribution for
$\dot{M}_{\rm{atm}} < m_{\rm{p}}/t_{\rm{sink}}$
(or equivalently for $\dot{M}_{\rm{atm}}/\dot{M}_{\rm{in}} < n^{-1}$) as
\begin{equation}
  P(\dot{M}_{\rm{atm}}) = \left( \frac{t_{\rm{sink}}}{m_{\rm{p}}} \right)^n
                          \frac{{\rm{e}}^{-n\gamma}}{\Gamma(n)}
                          \dot{M}_{\rm{atm}}^{n-1},
\end{equation}
where $\gamma \approx 0.577215665$ is Euler's constant and $\Gamma(n)$ is the gamma
function (see eq.~\ref{eq:a6}).
This means that the cumulative density distribution is
\begin{equation}
  f(>\dot{M}_{\rm{atm}}) = 1 - \frac{{\rm{e}}^{-n\gamma}}{n\Gamma(n)}
                           \left( \frac{\dot{M}_{\rm{atm}}t_{\rm{sink}}}
                                       {m_{\rm{p}}} \right)^n.
\end{equation}

\begin{figure}
  \begin{center}
    \vspace{-0.1in}
    \begin{tabular}{c}
      \hspace{-0.3in} 
      \psfig{figure=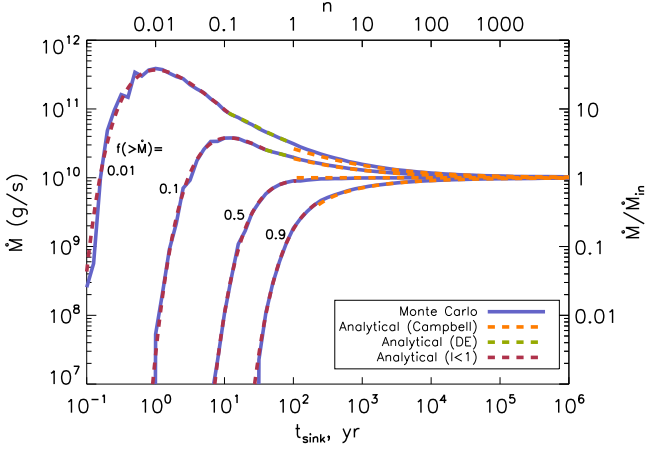,height=2.4in}
    \end{tabular}
    \caption{Simulations of accretion of $3.2 \times 10^{19}$ g planetesimals
    at a mean rate $10^{10}$ g\,s$^{-1}$ onto a white dwarf with a sinking time
    $t_{\rm{sink}}$.
    The lines show the distribution of inferred accretion rates;
    e.g., the top line corresponds to the level that would be exceeded in 1\% of
    measurements, while the $f(>\dot{M}_{\rm{atm}})=0.5$ line is the median
    of the distribution.  
    The blue solid line shows the results of an expanded set of Monte Carlo simulations
    similar to those shown in Fig.~\ref{fig:mcmono}.
    The dashed lines show various analytical estimates discussed in the text: 
    the $\dot{M}_{\rm{atm}}<m_{\rm{p}}/t_{\rm{sink}}$ solution in purple,
    the solution to the Gilbert \& Pollack (1960) differential difference equation
    in green,
    and Campbell's theorem in orange.
    The top and right axes generalise this plot to dimensionless accretion
    rate ($\dot{M}/\dot{M}_{\rm{in}}$) as a function of number of accretion
    events per sinking time ($n$).
    }
   \label{fig:analytmono}
  \end{center}
\end{figure}

Rather than compare this prediction directly with the distribution derived from the
Monte Carlo model in Fig.~\ref{fig:mcmono}c, we instead use those distributions to find
the 1\%, 10\%, 50\% and 90\% points in the distribution, repeat for a larger
number of sinking times, and plot these as a function of $t_{\rm{sink}}$ in
Fig.~\ref{fig:analytmono}.
Abbreviating $f(>\dot{M}_{\rm{atm}})$ to $f$ for now, the prediction is that
\begin{equation}
  \dot{M}_{\rm{atm}}(f) = \left( \frac{m_{\rm{p}}}{t_{\rm{sink}}} \right)
                          [(1-f)n\Gamma(n){\rm{e}}^{n\gamma}]^{1/n},
\end{equation}
which will be valid as long as the quantity in square brackets is less than 1
(e.g., for $n=1$, i.e. $t_{\rm{sink}}=100$~yr, this is valid for $f>1-\rm{e}^{-\gamma}
\approx 0.44$).
This is plotted in purple on Fig.~\ref{fig:analytmono} showing excellent agreement
with the Monte Carlo model, noting that deviations from the analytical
prediction are expected due to small number statistics.

For heuristic purposes, it is also worth pointing out that the distributions in the
limit of $n \ll 1$ for $f(>\dot{M}_{\rm{atm}}) \ll 1$ are asymptotically the same
as would be expected had we imagined planetesimals to arrive at regularly spaced
intervals of $t_{\rm{sink}}/n$ in time.
In that case, the fraction of time we would expect to infer accretion rates of
different levels would be determined by the exponential decay, and so
\begin{equation}
  f(>\dot{M}_{\rm{atm}}) = n \ln{ \left[ \frac{m_{\rm{p}}}
                            {\dot{M}_{\rm{atm}}t_{\rm{sink}}} \right] }
\end{equation}
in the range 1 to $\rm{e}^{-1/n}$ times $m_{\rm{p}}/t_{\rm{sink}}$.

There is no exact solution for the distribution at higher accretion rates
($\dot{M}_{\rm{atm}} > m_{\rm{p}}/t_{\rm{sink}}$), however
Gilbert \& Pollack provide a differential difference equation that can be solved to
determine $P(\dot{M}_{\rm{atm}})$ (see eq.~\ref{eq:a5}).
We show the resulting solution in green on Fig.~\ref{fig:analytmono}, but only over a
limited region of parameter space as validation of the technique, and of the Monte
Carlo model, since these are essentially different numerical methods of obtaining the
same answer.

However, there is an asymptotic solution in the large $n$ regime
(i.e., large $t_{\rm{sink}}$).
Campbell's theorem (Campbell 1909) can be applied to show that the probability
density function in this limit becomes a Gaussian with a mean of
$\dot{M}_{\rm{in}}$ (see eqs.~\ref{eq:a8} and \ref{eq:a9})
\begin{equation}
  P(\dot{M}_{\rm{atm}}) = \left( \frac{t_{\rm{sink}}}{m_{\rm{p}}} \right)
                          \frac{1}{\sqrt{2 \pi \sigma^2}}
          \rm{e}^{-\frac{1}{2\sigma^2}(\dot{M}_{\rm{atm}}-\dot{M}_{\rm{in}})^2},
\end{equation}
where the variance $\sigma^2=n(m_{\rm{p}}/t_{\rm{sink}})^2/2$.
This means that the cumulative distribution function is
\begin{equation}
  f(>\dot{M}_{\rm{atm}}) = [1-\rm{erf}(x)]/2,
  \label{eq:ferf}
\end{equation}
where $\rm{erf}(x)$ is the error function of
$x=\frac{\dot{M}_{\rm{atm}}-\dot{M}_{\rm{in}}}{\sqrt{\dot{M}_{\rm{in}}m_{\rm{p}}/t_{\rm{sink}}}}$.
Equation~(\ref{eq:ferf}) can be solved to get the appropriate points in the
distribution shown in orange on Fig.~\ref{fig:analytmono} for $n>1$.
This shows that Campbell's theorem provides an adequate approximation for large $n$,
but that discrepancies become noticeable as $n$ approaches 1.

\section{Can a mono-mass planetesimal distribution fit the observations?}
\label{s:mono}
It is clear from \S \ref{s:mod} that even with a very simple model, in which
planetesimals have the same mass around all stars, and in which all stars are
accreting matter at the same mean rate, it is expected that a broad distribution
of accretion rates could be inferred observationally, and that this distribution
could be different toward white dwarfs with different sinking times.
However, in \S \ref{ss:monomono} we explain why such a simple model
cannot explain the observationally inferred rates of \S \ref{s:obs}.
Then in \S \ref{ss:monomulti} we explore the possibility that all stars have
planetesimals that are the same mass, but that different stars have different mean
accretion rates, again ruling this out.
In \S \ref{ss:tdisc}, we consider how these conclusions may be affected if
planetesimals are processed through a disc on a timescale that can exceed the
sinking timescale before being accreted.

Throughout the paper we quantify the goodness-of-fit for a model in a
given sinking time bin $s$ as
\begin{equation}
  \chi_s^2 = \sum_j \left( \frac{f(>\dot{M}_{{\rm{obs}}(j,s)}) - f(>\dot{M}_{{\rm{atm}}(j,s)})}
                                {\sigma[f(>\dot{M}_{{\rm{obs}}(j,s)})]} \right)^2, \label{eq:chisq}
\end{equation}
where $f(>\dot{M}_{{\rm{obs}}(j,s)})$ is the best estimate from the observations of the
fraction of stars in bin $s$ with accretion above a level denoted by the index $j$, where the
sum is performed for $j$ corresponding to $10^7$, $10^8$, $10^9$ and $10^{10}$~g\,s$^{-1}$,
$\sigma[f(>\dot{M}_{{\rm{obs}}(j,s)})]$ is the larger of the positive or negative
$1\sigma$ uncertainties plotted on Fig.~\ref{fig:obs}d,
and $f(>\dot{M}_{{\rm{atm}}(j,s)})$ is the corresponding model distribution.
Since the observables in a cumulative distribution
(i.e., $f(>\dot{M}_{{\rm{obs}}(j,s)})$) are not independent at the
different indices, the absolute value of $\chi_s^2$ should not be used to 
determine the formal significance of the model fit to the data.
Rather we will be using it here as a relative measure of the goodness-of-fit
of different models for a given bin.

\subsection{Mono-mass, mono-rate accretion}
\label{ss:monomono}
The distribution of inferred accretion rates for a mono-mass mono-rate
model will always have a dependence on sinking time that is similar in form to that
shown in Fig.~\ref{fig:mcmono}c.
Varying the mean accretion rate parameter, $\dot{M}_{\rm{in}}$, would simply
change the x-axis scaling such that the distributions for the longest sinking
times all have accretion rates inferred at the $\dot{M}_{\rm{in}}$ level (see top
axis). 
Varying the planetesimal mass would change the sinking times corresponding to
the different lines on the figure, but these lines would always correspond to
the same $n$ given in the legend (e.g., the green line corresponds to $n=1$), and
so eq.~(\ref{eq:n}) can be used to work out the corresponding sinking time which
just scales with planetesimal mass (e.g., the pale green line corresponds to
$t_{\rm{sink}}=m_{\rm{p}}/\dot{M}_{\rm{in}}$). 

Fig.~\ref{fig:obs}d provides several clues as to what 
combination of $\dot{M}_{\rm{in}}$ and $m_{\rm{p}}$ would be required to reproduce any given
distribution.
For example, the fact that the $\dot{M}_{\rm{obs}}$ distribution is broad for all of the
sinking time bins means that $n \ll 1$ for all of the bins.
The breadth of the distribution is indicative of the $n$ required to fit any of the relevant
lines, and the appropriate value for the long sinking timescale bin can be inferred
readily from Fig.~\ref{fig:analytmono} using the top and right axes.
That is, for there to be a range of around 1000 in accretion rates between the
10\% and 50\% points in the distribution requires $n \approx 0.09$.
The input accretion rate can then be found by scaling the 50\% point to be close to
$10^6$~g\,s$^{-1}$ (Fig.~\ref{fig:obs}d) giving an $\dot{M}_{\rm{in}}$ of around
$1.7 \times 10^8$~g\,s$^{-1}$, and so a planetesimal mass $m_{\rm{p}}$ of around
$2.2 \times 10^{22}$~g (for the median sinking time of 0.37~Myr in this bin).

\begin{figure}
  \begin{center}
    \vspace{-0.1in}
    \begin{tabular}{c}
      \hspace{-0.3in} 
      \psfig{figure=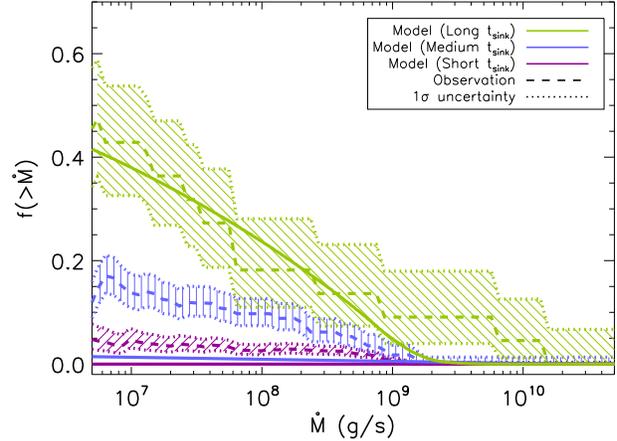,height=2.6in}
    \end{tabular}
    \caption{Simulations of accretion of $2.2 \times 10^{22}$~g planetesimals
    at a mean rate $1.7 \times 10^{8}$~g\,s$^{-1}$ onto populations of white dwarfs
    with distributions of sinking times that match that of the corresponding
    observed populations in each of the sinking time bins.
    The dashed lines show the distribution inferred from the observations, while
    the hatched regions and dotted lines show the $\pm 1\sigma$ range of possible
    distributions given small number statistics (reproduced from
    Fig.~\ref{fig:obs}d).
    The model predictions are shown with solid lines in the corresponding colour.
    The model for the short sinking time bin is indistinguishable
    from 0 on this plot.
    }
   \label{fig:monomono}
  \end{center}
\end{figure}

Fig.~\ref{fig:monomono} reproduces the inferred accretion rate
distributions of Fig.~\ref{fig:obs}d and also makes predictions for model
populations in which stars have the same distributions of sinking times as that 
of the observed population in the corresponding bin, under the assumption that all
stars are accreting $2.2 \times 10^{22}$~g planetesimals (i.e., roughly 240~km diameter
asteroids) at a mean rate $1.7 \times 10^8$~g\,s$^{-1}$ (equivalent to around 1 asteroid
belt every 680~Myr).
This model population was implemented by taking each star in the corresponding observed
population and running the Monte Carlo model of \S \ref{ss:monte} with the sinking time
for that star, then combining the results for all stars into one single population.
The number of timesteps used for each star, $N_{\rm{tot}}$, was chosen so that the total
number of accretion rates used for the model population (i.e., $N_{\rm{tot}}$ times the
number of stars in the observed population) was close to $10^5$.

As expected from the arguments two paragraphs ago, a model with these parameters
gives a decent fit to the long sinking time bin (for reference $\chi_s^2=1.0$
as defined in eq.~\ref{eq:chisq}).
However, the same model provides a very poor fit to the shorter sinking time bins
($\chi_s^2=12$ and 21 in the short and medium sinking time bins, respectively).
The problem is that having $n\ll1$ in the long bin means that such timescales are
already sampling the vestiges of past events (i.e., such events happen
much less frequently than once per Myr).
This means that, while it is possible for measurements with shorter sinking times
to infer high accretion rates just after the event, such measurements would be
extremely rare.
By consequence we would expect to see essentially no accretion signatures
in the samples with $t_{\rm{sink}}<0.1$~Myr (see Fig.~\ref{fig:monomono}).

\subsection{Mono-mass, multi-rate accretion}
\label{ss:monomulti}
One conclusion from \S \ref{ss:monomono} is that, for a mono-mass
distribution of planetesimal masses, a model that fits all sinking
time bins simultaneously requires $n \gg 1$ for (the majority of) the long
sinking time bin.
The broad distribution of observationally inferred accretion rates in this bin,
$f(>\dot{M}_{\rm{obs}})$, thus implies that different stars accrete
at different rates, and that the observationally inferred distribution is representative
of that of the mean rate at which material is being accreted,
$f(>\dot{M}_{\rm{in}})$.
At least this must be the case for high accretion rates, but it is possible that
the lowest accretion rates, say below $\dot{M}_{\rm{in}}=10^7$~g\,s$^{-1}$,
are in the $n<1$ regime.
This also sets a constraint on the planetesimal mass, since requiring
$n>1$ in the $\sim 0.37$~Myr sinking time bin for $\sim 10^7$~g\,s$^{-1}$ means that
$m_{\rm{p}}<10^{20}$~g.

Here we modify the population model of \S \ref{ss:monomono} by assuming
that different white dwarfs accrete at different rates, i.e. that there is a distribution
$f(>\dot{M}_{\rm{in}})$, but from the same mono-mass distribution of planetesimal masses,
$m_{\rm{p}}$.
Practically this is implemented in the model population by each of the observed stars in
the appropriate sample having its accretion rate chosen randomly from
the given distribution a sufficient number of times to get a total of $\sim 1000$ combinations
of $t_{\rm{sink}}$ and $\dot{M}_{\rm{in}}$, which are then simulated at 1000 timesteps.
For an assumed planetesimal mass, we proceed by using
the long sinking time bin to constrain the distribution
$f(>\dot{M}_{\rm{in}})$.
Comparison of the model predictions to the observationally inferred rates
for all sinking time bins
is then used to determine the planetesimal mass that gives the best overall fit.

The simplest form for the distribution of $\dot{M}_{\rm{in}}$
is log-normal, with a median of $10^{\mu}$~g\,s$^{-1}$ and width of $\sigma$~dex.
As pointed out above, if the planetesimal mass is small enough this distribution
should be defined by the distribution of rates inferred in the long sinking time bin.
By minimising $\chi_s^2$ for the long bin, the median and the width of the observationally
inferred distribution were found to be $\mu=6.6$ and $\sigma=1.5$, and we confirmed that using
this for the input accretion rate, $f(>\dot{M}_{\rm{in}})$, gives a reasonable fit
to the long sinking time bin provided that $m_{\rm{p}}<10^{20}$~g.

\begin{figure}
  \begin{center}
    \vspace{-0.1in}
    \begin{tabular}{cc}
      \hspace{-0.1in} \textbf{(a)} &
      \hspace{-0.3in} 
      \psfig{figure=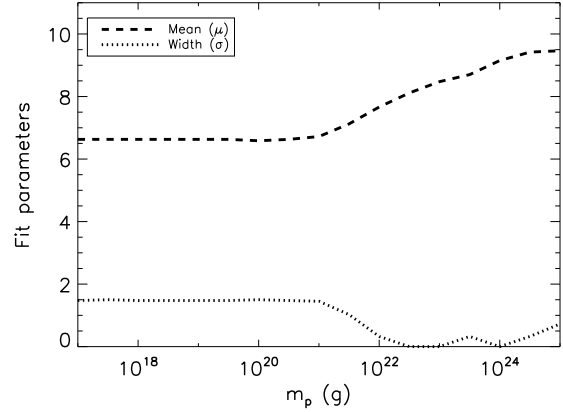,height=2.4in} \\[-0.0in]
      \hspace{-0.1in} \textbf{(b)} &
      \hspace{-0.3in} 
      \psfig{figure=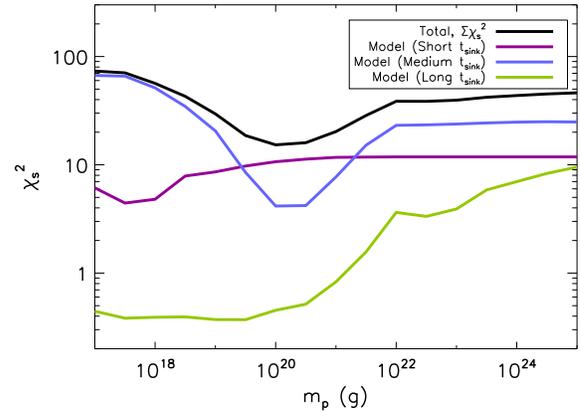,height=2.4in} \\[-0.0in]
      \hspace{-0.1in} \textbf{(c)} &
      \hspace{-0.3in} 
      \psfig{figure=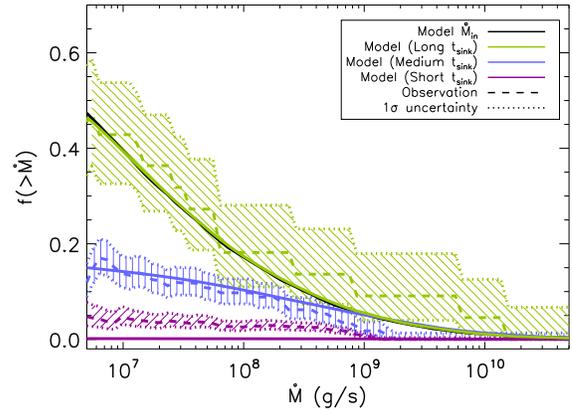,height=2.4in} \\
    \end{tabular}
    \caption{Simulations of accretion of planetesimals all of mass $m_{\rm{p}}$,
    at a mean rate drawn from a log-normal distribution described by the parameters $\mu$ and $\sigma$
    for populations with the same distribution of sinking times as the stars observed in
    the corresponding bins in Fig.~\ref{fig:obs}d.
    \textbf{(a)} Parameters for the input accretion rate distribution that give a best
    fit to the long sinking time bin.
    \textbf{(b)} The goodness of fit $\chi_s^2$ to the 3 different bins as a function of $m_{\rm{p}}$.
    \textbf{(c)} Comparison of the model populations to the rates inferred from the observations
    for the parameters providing the best (but still not great) fit to all bins, which is for
    $m_{\rm{p}}=1.0 \times 10^{20}$~g (see \textbf{(b)}).
    The model for the short sinking time bin is indistinguishable
    from 0 on this plot.
    }
   \label{fig:monomulti}
  \end{center}
\end{figure}

As $m_{\rm{p}}$ is increased above $10^{20}$~g, the input accretion rate distribution given in
the last paragraph no longer provides a reasonable fit to the long sinking time bin, as a larger
fraction of stars in the sample have $n<1$.
To get around this the input accretion rate needs to be higher (because the
inferred rate is lower for most stars when $n<1$) and the width of the
distribution narrower (since decreasing $n$ leads to a broader distribution of
inferred accretion rates).
The parameters of a log-normal input accretion rate distribution that give the
minimum $\chi_s^2$ for the long sinking time bin are given in Fig.~\ref{fig:monomulti}a
as a function of assumed planetesimal mass.
As can be seen, these tend to the values given in the last paragraph for small $m_{\rm{p}}$,
and change in the sense expected as $m_{\rm{p}}$ is increased.

Fig.~\ref{fig:monomulti}b shows how the resulting goodness of fit $\chi_s^2$ varies for all
sinking time bins as $m_{\rm{p}}$ is changed.
This shows that it is not possible to maintain a reasonable quality fit to the long
bin with high $m_{\rm{p}}$.
This is inevitable, because in the regime of large $m_{\rm{p}}$ (i.e., small $n$),
the distribution of inferred accretion rates necessarily becomes very broad even for
input distributions that are very narrow (see Figs.~\ref{fig:mcmono}c and
\ref{fig:analytmono}), and eventually become much broader than that inferred from the
observations.
Thus the best fit will tend to one in which the model has too many high accretion rates, but
too few low accretion rates.
It is no coincidence that the best fit to this bin starts to get significantly worse beyond
around $m_{\rm{p}}=2 \times 10^{22}$~g at a point close to $\mu=8.2$ and $\sigma=0$, which was the
best fit of \S \ref{ss:monomono}.

Even for planetesimal masses where a reasonable fit to the long sinking time bin is possible,
it is not possible to simultaneously find an acceptable fit to both shorter sinking time bins.
Fig.~\ref{fig:monomulti}b shows how $\chi_s^2$ varies with planetesimal mass, and
Fig.~\ref{fig:monomulti}c shows the best fit that minimises the sum of $\chi_s^2$ for all bins,
which is for $m_{\rm{p}}=1.0 \times 10^{20}$~g.
For example, consider the medium sinking time bin.
For the smallest planetesimal masses, $n \gg 1$ for all stars in this bin and so the distribution of
inferred rates for the model population is close to that for the input rates;
i.e., the model population has too many large accretion rates.
Increasing planetesimal mass decreases $n$ for all stars, and a crude approximation
is that the resulting distribution remains close to that of the input rates for large
accretion rates, but becomes flat at accretion rates for which 
$n \ll 1$ for the bin's median sinking time of 850~yr, corresponding to
$\dot{M} \ll 3.7 \times 10^{9}$~g\,s$^{-1}$ on Fig.~\ref{fig:monomulti}c (compare the black and blue lines);
a similar analysis for the long sinking time bin explains why the model's inferred
accretion rate distribution (the green line on Fig.~\ref{fig:monomulti}c) only departs from the input rate
distribution (black line) below $9 \times 10^{6}$~g\,s$^{-1}$.
Increasing planetesimal mass above the best fit of $1.0 \times 10^{20}$~g
results in the model population having a negligible
fraction with accretion rates in the appropriate range.
The situation is similar for the short sinking time bin
(see Fig.~\ref{fig:monomulti}b), except that the model population
is closest to that inferred from the observations
(albeit slightly flat) when the planetesimal mass is
just below $10^{18}$~g, with essentially no accretion signatures expected in this bin by
the time the planetesimal mass is large enough to fit the medium bin
(Fig.~\ref{fig:monomulti}c). 

In conclusion, although the best fit has improved relative to \S \ref{ss:monomono},
it is not possible to provide a reasonable fit to the 
distribution of accretion rates inferred from the observations within the constraints of this model.
In \S \ref{ss:tdisc} we explore whether this is primarily an (avoidable) consequence of
having so many orders of magnitude difference in sinking times between the bins.

\subsection{Finite disc lifetime}
\label{ss:tdisc}
The assumption thus far is that the entire planetesimal mass is 
placed in the stellar atmosphere on a timescale that is much shorter than the
sinking timescale. 
If material were accreted by direct impact onto the star this would be
reasonable.
However, accretion by direct impact is considered unlikely since the stellar radius is
so much smaller ($\sim 70$ times) than the tidal disruption radius, meaning that
material is likely to tidally disrupt and form a disc before whatever process that
kicked it to $1R_\odot$ gets it onto the star (Farihi et al. 2012b).
Indeed observations support the notion that material is processed through
a disc that is sometimes detectable (see \S \ref{s:intro}).
The lifetime of such discs and the physical mechanisms by which material is accreted
onto the star are active topics of discussion (Rafikov 2011; Metzger et al. 2012; Farihi
et al. 2012b).
Various timescales are involved, such as that to circularise
the orbits of tidally disrupted planetesimals, that to convert this material into 
dust, that to make the dust reach the sublimation radius where it is converted
into gas, and that for gas to accrete onto the star.
Nevertheless, it would not be unreasonable to assume that this process takes many years,
since the viscous time for gas to get from the sublimation radius to the star is at
least 100-1000 years (Metzger et al. 2012; Farihi et al. 2012b).

To model the disc properly requires significant modifications to the model that are
beyond the scope of this paper.
Instead the disc will be considered here with the simplest prescription that is readily
implemented into the model.
Thus we assume that all discs have the same lifetime $t_{\rm{disc}}$, and that 
following accretion (by which we really mean incorporation into the disc), the
planetesimal mass present in the atmosphere decays on a timescale
\begin{equation}
  t_{\rm{samp}} = \sqrt{t_{\rm{disc}}^2 + t_{\rm{sink}}^2}.
  \label{eq:tsamp}
\end{equation}
Effectively this means that, even if a star has a sinking time of a few days, the
timescale over which observations of the atmospheric pollution are sampling
the accretion rate can be longer, and this timescale is roughly equal to the larger
of the disc lifetime and the sinking time.
The question then is whether this additional parameter is sufficient to allow us to fit
the distributions inferred from the observations, and if so what is the typical disc lifetime.

\subsubsection{Mono-mass, mono-rate with disc lifetime}
\label{sss:monomonotdisc}

\begin{figure}
  \begin{center}
    \vspace{-0.1in}
    \begin{tabular}{cc}
      \hspace{-0.1in} \textbf{(a)} &
      \hspace{-0.3in} 
      \psfig{figure=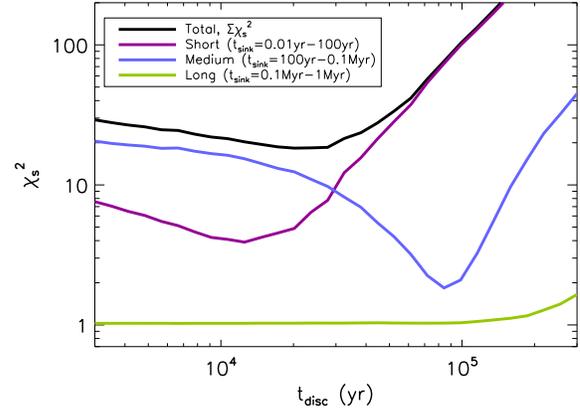,height=2.4in} \\[-0.1in]
      \hspace{-0.1in} \textbf{(b)} &
      \hspace{-0.3in} 
      \psfig{figure=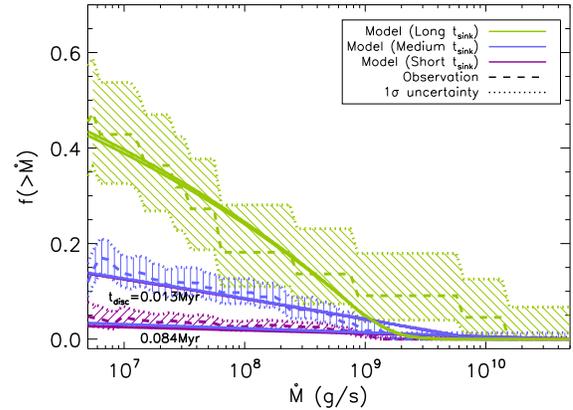,height=2.4in} \\[-0.0in]
    \end{tabular}
    \caption{Simulations of accretion of $2.2 \times 10^{22}$~g planetesimals
    at a mean rate $1.7 \times 10^{8}$~g\,s$^{-1}$ that are identical to those for
    Fig.~\ref{fig:monomono}, except that sampling times combine
    both the sinking time in the white dwarf atmosphere and a disc lifetime ($t_{\rm{disc}}$).
    \textbf{(a)} Goodness-of-fit $\chi_s^2$ as a function of disc lifetime for the different
    sinking time bins, as well as for all bins combined.
    \textbf{(b)} Comparison of the model populations to the rates inferred from observations for the
    disc lifetimes that provide the best fit to the short sinking time bin ($t_{\rm{disc}}=0.013$~Myr) and to
    the medium sinking time bin ($t_{\rm{disc}}=0.084$~Myr).
    For each of these disc lifetimes the model populations in the
    medium and short sinking time bins are very similar and so are hard to differentiate.
    The model populations for the long sinking time bins are indistinguishable for the
    two disc lifetimes.
    }
   \label{fig:monomono2}
  \end{center}
\end{figure}

In Fig.~\ref{fig:monomono2} we repeat the modelling of \S \ref{ss:monomono} to show
that a reasonable fit could be obtained simultaneously with both
the long bin and either the medium bin (with $t_{\rm{disc}} \approx 0.084$~Myr)
or the short bin (with $t_{\rm{disc}} \approx 0.013$~Myr), but that it is not possible
to fit all bins simultaneously.
The problem is that disc lifetimes of $>0.01$~Myr are so high that the sampling
times for the populations of both shorter timescale bins are set by the
disc lifetime and so are very similar.
Consequently, their inferred accretion rate distributions are indistinguishable.
For these bins to have different distributions requires $t_{\rm{disc}} \ll 0.01$~Myr,
but that results in too few stars having accretion rates in the range of those
inferred observationally.
Since the accretion rate distributions inferred observationally for these bins
differ at the $2-3\sigma$ level (\S \ref{ss:distrn}), we consider that while this model with a disc lifetime
in between $0.01$ and $0.08$~Myr would provide a reasonable fit to the observationally
inferred distributions (e.g., if the
medium and short sinking time bins were combined), this is mildly
disfavoured by the observations.
Thus we continue to try to find a model that also predicts a difference between the short and medium
sinking time bins.

\subsubsection{Mono-mass, multi-rate with disc lifetime}
\label{sss:monomultitdisc}
To repeat the modelling for the case that the input accretion rates are not necessarily
the same for all stars (i.e., modelling analogous to \S \ref{ss:monomulti}) it is helpful
to note that the long sinking time bin must be unaffected by the disc lifetime, because
otherwise the distributions for all sinking time bins would look the same.
This means that the $\mu$ and $\sigma$ of the distribution of input accretion rates
required to fit the long bin can be taken from Fig.~\ref{fig:monomulti}a.
Thus these parameters were fixed (for a given $m_{\rm{p}}$), and the modelling was repeated
for a range of $t_{\rm{disc}}$.
For each disc lifetime, $m_{\rm{p}}$ was chosen
so as to minimise $\chi_s^2$ either of each of the sinking time bins, or of the
total $\chi_s^2$ for all bins.

Some general comments can be made on how the best fit parameters and the resulting
accretion rate distributions vary with assumptions about disc lifetime.
For $t_{\rm{disc}} \ll 1$~yr the solution is unaffected by the disc lifetime, and
the solution tends to that of Fig.~\ref{fig:monomulti}.
For $t_{\rm{disc}} \gg 0.1$~Myr, on the other hand, all bins have the same sampling
time (i.e., close to the disc lifetime) and so all have the same distribution of
inferred accretion rates.
Since neither extreme provides a reasonable fit to the observationally inferred distributions, but for very
different reasons --- for small $t_{\rm{disc}}$ the distributions of the different bins are
too far apart, whereas for large $t_{\rm{disc}}$ the distributions are too similar ---
it might be hoped that an intermediate value of $t_{\rm{disc}}$ would improve the
fit.
This is indeed the case, however, the improvement is very small, since an intermediate disc
lifetime is the situation described in \S \ref{sss:monomonotdisc}, and the problems of that
model are not much ameliorated by allowing there to be a distribution of input accretion rates;
that is, it is still not possible to separate the three sinking time bins.

Thus we conclude that none of the mono-mass planetesimal distribution models provide
an adequate fit to the observationally inferred accretion rate distributions, with the
caveat that this requires those distributions for the three sinking time bins to be
different from each other, which needs confirmation.
The line-of-reasoning outlined above also suggests that if thermohaline convection modifies
the rates such that these are independent of sinking time (see \S \ref{ss:thermohaline}), this could be
used to argue for a disc lifetime $\gg 0.1$~Myr, in which case a mono-mass planetesimal distribution
remains a possibility.


\section{Models of accretion from planetesimals with a range of masses}
\label{s:mod2}
In this section we relax the assumption that the accreted material is all in
planetesimals that are of the same mass (i.e., mono-mass), and instead
assume that material is accreted at a mean rate $\dot{M}_{\rm{in}}$
from a power law mass distribution that is defined by the index $q$.
That is, if $n(m)dm$ is the number of objects in the mass range $m$ to $m+dm$
then 
\begin{equation}
  n(m) \propto m^{-q}, \label{eq:nd}
\end{equation}
where this parameterisation means that the commonly quoted index on the
size distribution (i.e., $n(D) \propto D^{-\alpha}$) would be $\alpha=3q-2$ for
spherical particles of constant density. 
If we assume that $q<2$, and that the most massive planetesimal
in the distribution, of mass $m_{\rm{max}}$, is much more massive than the
least massive dust grain, of mass $m_{\rm{min}}$,
then the majority of the mass is in the largest objects and the new model is
simply defined by two parameters ($q$ and $m_{\rm{max}}$) instead of
one ($m_{\rm{p}}$).
While one might imagine that this would be
equivalent to a mono-mass distribution with mass $m_{\rm{p}} \sim m_{\rm{max}}$,
this is not the case if the number of planetesimals with mass $m_{\rm{max}}$
arriving per sampling time (i.e., the larger of the sinking time and disc lifetime,
eq.~\ref{eq:tsamp}) is less than unity.
In \S \ref{ss:params} we show how the distribution of accretion rates that would be inferred
is more closely related to planetesimals of mass $m_{\rm{tr}} < m_{\rm{max}}$ for which
the total number of planetesimals with masses larger than $m_{\rm{tr}}$ arriving each
sampling time is of order unity.
Then in \S \ref{ss:representative} we show that this model can be used to provide a
reasonable fit to the observationally inferred distribution of accretion rates.

\subsection{Simple model}
\label{ss:params}
\begin{figure}
  \begin{center}
    \vspace{-0.1in}
    \begin{tabular}{cc}
      \hspace{-0.2in} \textbf{(a)} & \hspace{-0.2in} 
      \psfig{figure=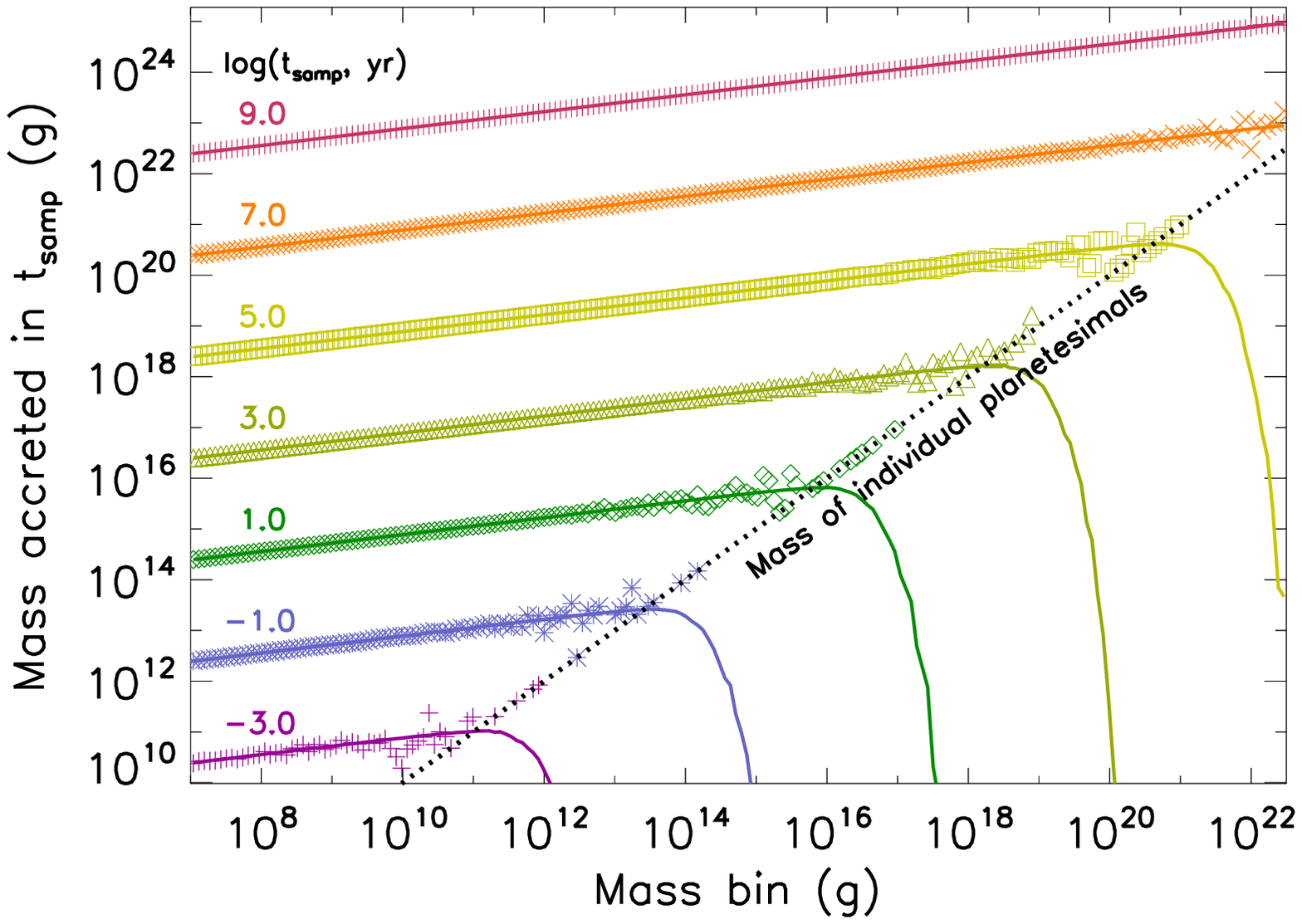,height=2.4in} \\[0.1in]
      \hspace{-0.2in} \textbf{(b)} & \hspace{-0.2in} 
      \psfig{figure=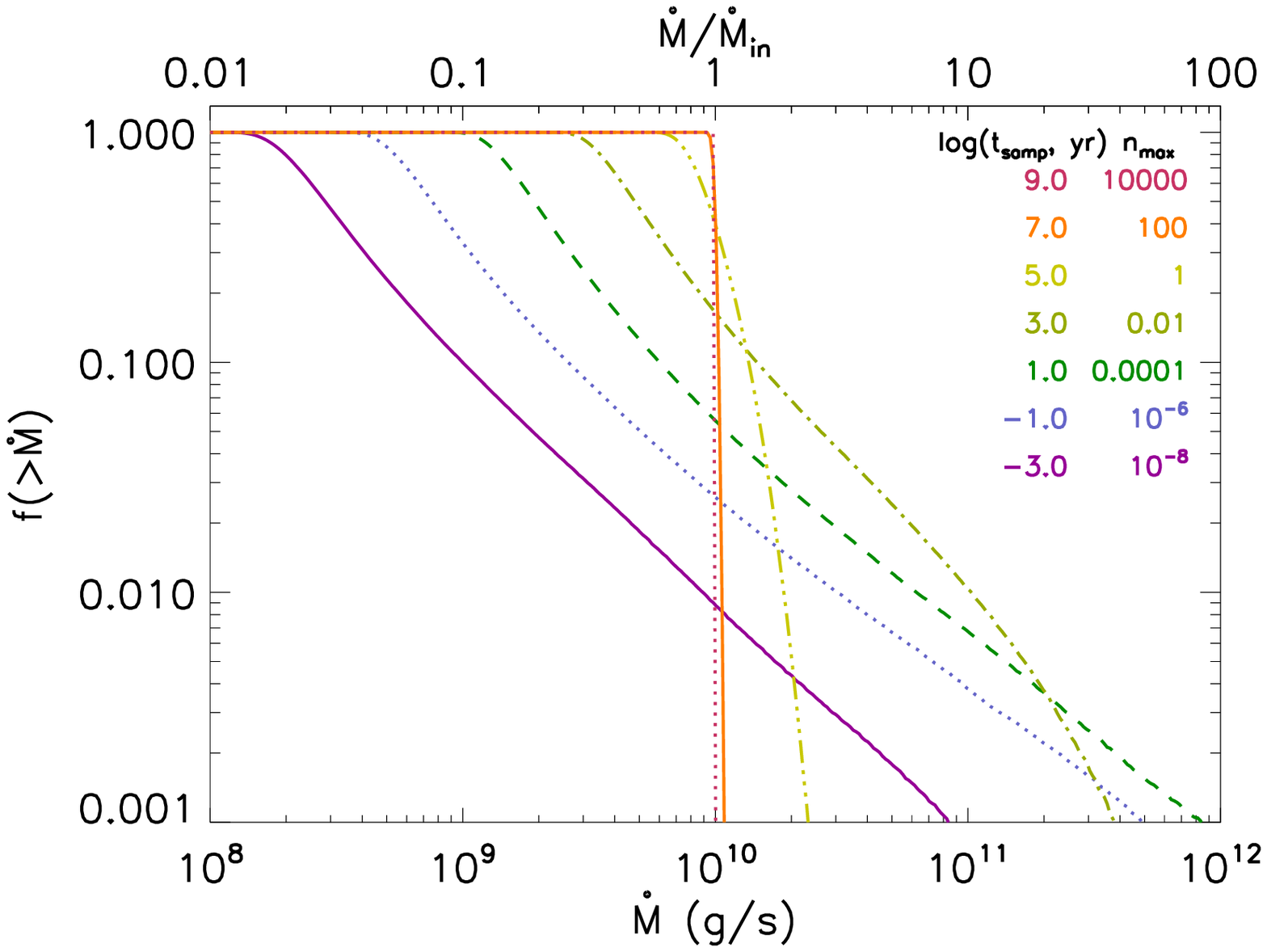,height=2.4in} 
    \end{tabular}
    \caption{Monte carlo simulations of sampling a mass distribution with
    $m_{\rm{max}}=3.2 \times 10^{22}$~g and a power law index $q=11/6$ at a rate
    $10^{10}$~g\,s$^{-1}$ with different sampling times $t_{\rm{samp}}$.
    \textbf{(a)} Snapshot of the mass accreted from different logarithmically spaced
    mass bins over a time interval of one sampling time (symbols), where models with different
    sampling times are shown with different colours. 
    The solid coloured lines use the results of many snapshots to show the median mass remaining
    in the atmosphere from the accretion of material from this bin.
    \textbf{(b)} Distribution of accretion rates that would be inferred after many realisations of
    the snapshots seen in \textbf{(a)}.
    }
   \label{fig:model3.5}
  \end{center}
\end{figure}

To illustrate the effect of planetesimals having a distribution of masses,
Fig.~\ref{fig:model3.5} shows the predictions of a Monte Carlo model of
accretion from a distribution in which $q=11/6$ and $m_{\rm{max}}=3.16 \times 10^{22}$~g.
To do this, $N_{\rm{m}}=200$ logarithmically spaced mass bins
were set up down to an inconsequentially small minimum
mass of $m_{\rm{min}}=10^{7}$~g.
The logarithmic width of the bin is
$\delta=(\log{m_{\rm{max}}}-\log{m_{\rm{min}}})/N_{\rm{m}}$, and bins are referred
to by their index $k$, so that planetesimals in the bin have a typical mass denoted $m_k$.
Assuming a mean accretion rate of $\dot{M}_{\rm{in}} = 10^{10}$~g\,s$^{-1}$,
the amount of mass accreted from each bin in a given time interval, and the amount
of mass that remains in the atmosphere from previous accretion events from that bin
(for a given sampling time $t_{\rm{samp}}$),
was then modelled in exactly the same way as described for the accretion of 
a mono-mass planetesimal distribution (\S \ref{ss:monte}).
The results for all of the bins were then combined to get the expected
distribution of mass in the atmosphere for the $N_{\rm{tot}} = 200,000$ timesteps.
This process was repeated for different sampling times in the range
$t_{\rm{samp}}=10^{-3}$ to $10^{9}$~yr.

The snapshot shown in Fig.~\ref{fig:model3.5}a illustrates how the accretion
from different mass bins can be divided into a continuous and a stochastic
component.
For a given sampling time, for small enough planetesimal masses, the mass accreted
from different bins simply follows the mass distribution, with mass accreted in
the interval of duration $t_{\rm{samp}}$ being
\begin{equation}
  M_{\rm{ac}(k)} = (2-q) \dot{M}_{\rm{in}} t_{\rm{samp}}
                     \delta (m_k/m_{\rm{max}})^{2-q}.
  \label{eq:msampjk}
\end{equation}
However, since only integer numbers of particles can be accreted in any one timestep,
this relation breaks down for bins for which the mass that would have been
expected to be accreted is comparable with that of a single planetesimal.

As noted in \S \ref{s:mod}, what is important is the mean number
of planetesimals accreted from bin $k$ per sampling time, $M_{\rm{ac}(k)}/m_k$.
However, to avoid having model parameters that depend on bin size $\delta$, here
we integrate equation (\ref{eq:msampjk}) from $m_{\rm{min}}$ to $m_k$ to get the mass
accreted in $t_{\rm{samp}}$ from objects smaller than $m_k$.
We then use this to work out the number of planetesimals of mass $m_k$ that would need
to be accreted per sampling time to maintain that accretion rate
\begin{eqnarray}
  n_k & = & n_{\rm{max}} (m_k / m_{\rm{max}})^{1-q}, \label{eq:nk} \\
  n_{\rm{max}} & = & \dot{M}_{\rm{in}} t_{\rm{samp}} / m_{\rm{max}}, \label{eq:nmax}
\end{eqnarray}
where $n_{\rm{max}}$ is the mean number of the largest planetesimals in the distribution
that would need to be accreted to maintain the input accretion rate (if only those
largest planetesimals were present).

The planetesimal mass at which $n_k=1$, which we call 
\begin{equation}
  m_{\rm{tr}} = m_{\rm{max}} n_{\rm{max}}^{1/(q-1)},
  \label{eq:dtr}
\end{equation}
is that at which the mass accreted in $t_{\rm{samp}}$
from planetesimals less massive than $m_{\rm{tr}}$
is equal to a single planetesimal of mass $m_{\rm{tr}}$.
This mass marks the transition from continuous to stochastic accretion;
in any given timestep, most bins above $m_{\rm{tr}}$ would be expected to
have no planetesimals accreted from them, with the occasional bin offering up the accretion
of a single planetesimal.
The solid lines on Fig.~\ref{fig:model3.5}a show that the mass left in the atmosphere
from such bins is, in an average timestep, very small.
Thus the typically inferred accretion rate is dominated by the accretion of
objects of mass around $m_{\rm{tr}}$.

Fig.~\ref{fig:model3.5}b shows the distribution of mass accretion rates that would be
expected to be inferred, given the mass that would remain in the atmosphere for the
given sampling times, for the $N_{\rm{tot}}$ realisations of the model.
For long enough sampling times all planetesimal masses are accreted continuously and all
timesteps measure an accretion rate equal to the mean rate of $10^{10}$~g\,s$^{-1}$.
As a stochastic element only arises if $m_{\rm{tr}} < m_{\rm{max}}$, 
the sampling time above which stochasticity is unimportant can be estimated by
setting $m_{\rm{tr}}=m_{\rm{max}}$ in eq.~\ref{eq:dtr}, so that 
\begin{equation}
  t_{\rm{samp,crit}} = m_{\rm{max}}/\dot{M}_{\rm{in}};
\end{equation}
i.e., we would expect $t_{\rm{samp,crit}}$ to be 0.1~Myr for the parameters
given here, in agreement with Fig.~\ref{fig:model3.5}b for which the accretion rates
are in a narrow distribution around $10^{10}$~g\,s$^{-1}$ for
$\log{(t_{\rm{samp}})}\gg5$.

For sampling times significantly below this value, however, we expect different timesteps
to measure different accretion rates, depending on whether stochastic processes happen
to have favoured the timestep (or those in the recent past) with many or few objects of mass
around $m_{\rm{tr}}$ and above.
As mentioned previously, the distribution must still have a mean
of $\dot{M}_{\rm{in}}$.
However, for short sampling times the mean would be dominated by
events so rare (like the accretion of a planetesimal of mass $m_{\rm{max}}$) that it
is unlikely to be measured in any of our timesteps for a realistic value of $N_{\rm{tot}}$.
Nevertheless, our realisations give an indication of the median of the distribution,
and so of the typical level of accretion that would be seen.
It is notable that the median tends to smaller values for smaller sampling times.

\begin{figure}
  \begin{center}
    \vspace{-0.1in}
    \begin{tabular}{cc}
      \hspace{-0.2in} \textbf{(a)} & \hspace{-0.3in} 
      \psfig{figure=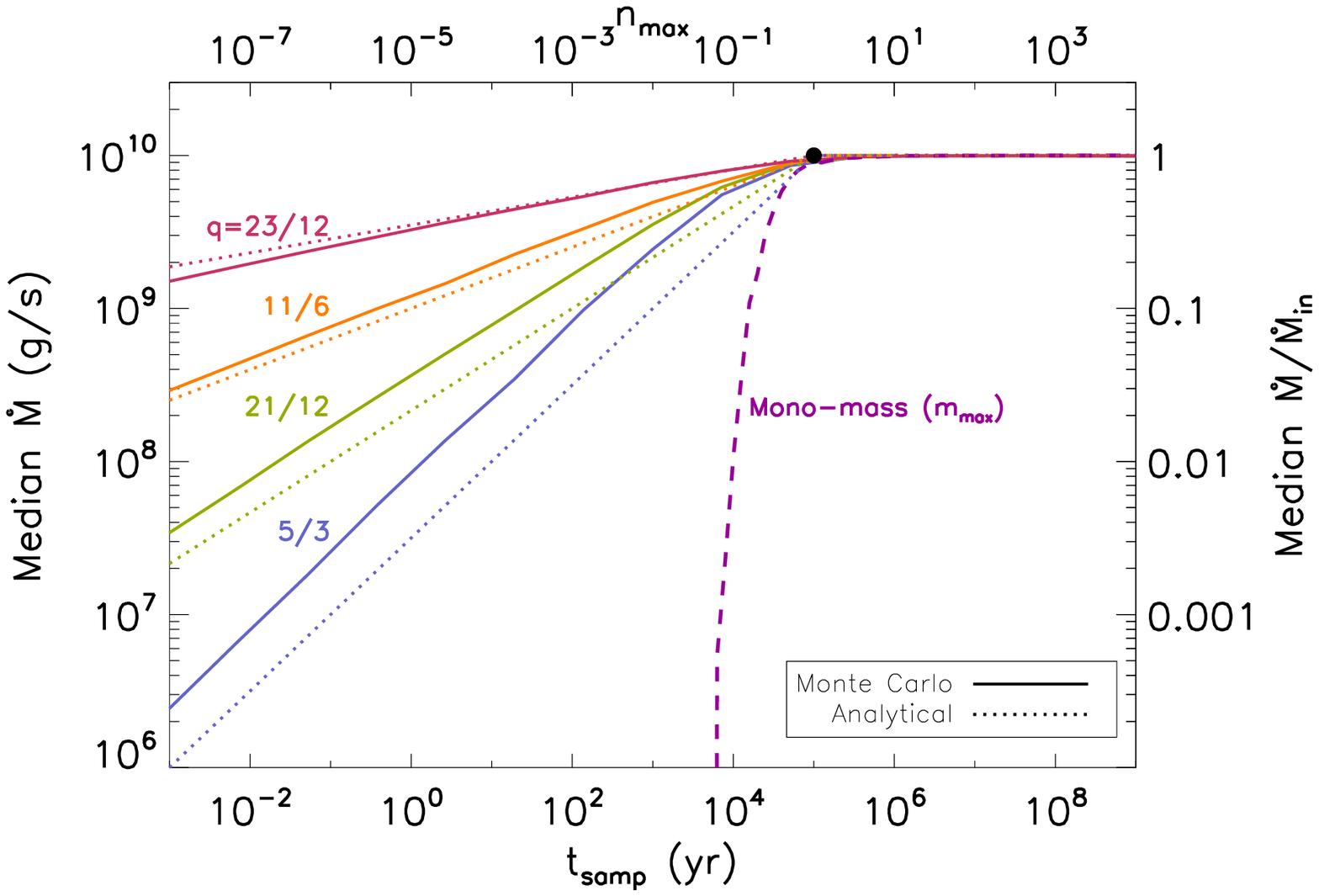,height=2.2in} \\
      \hspace{-0.2in} \textbf{(b)} & \hspace{-0.3in} 
      \psfig{figure=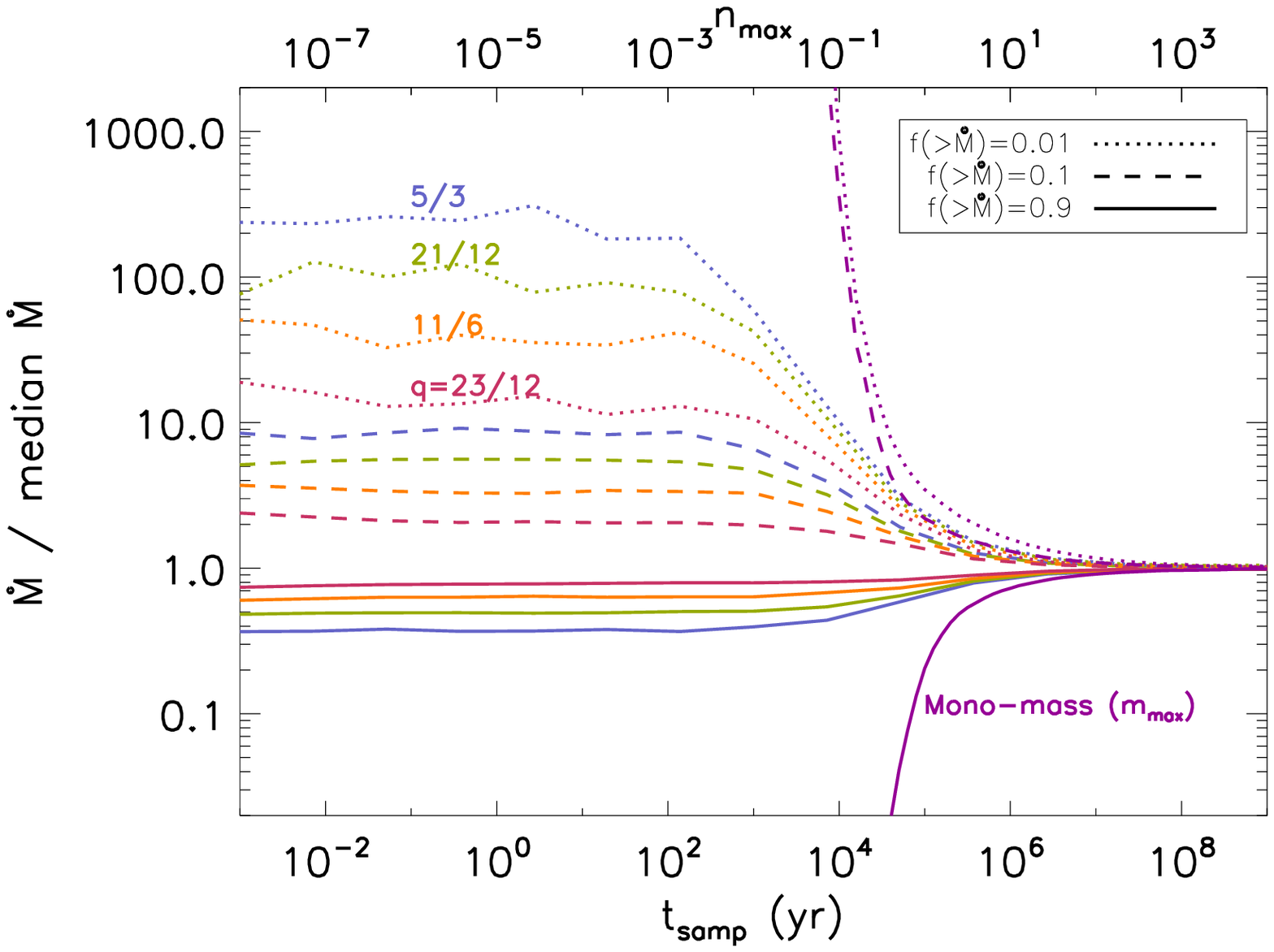,height=2.2in}
    \end{tabular}
    \caption{Monte carlo simulations of sampling a mass distribution with
    $m_{\rm{max}}=3.2 \times 10^{22}$~g and a power law index $q$ at a
    rate $10^{10}$~g\,s$^{-1}$ with different sampling times $t_{\rm{samp}}$.
    \textbf{(a)} Median accretion rate as a function of
    $t_{\rm{samp}}$ for models with different mass distribution slopes $q$
    (indicated with different colours).
    The solid line is the result of the Monte Carlo models, and the dotted line
    is the analytical prediction of eq.~(\ref{eq:mmed}).
    The mono-mass model of Fig.~\ref{fig:analytmono} (appropriately scaled so that all
    planetesimals are of mass $m_{\rm{max}}$) is shown with a dashed line.
    \textbf{(b)} The width of the distribution of accretion rates, as defined by the
    accretion rates for which 90\%, 10\% and 1\% of measurements are expected to have
    values higher than this (relative to the median accretion rate).
    }
   \label{fig:width}
  \end{center}
\end{figure}

To quantify the median accretion rate discussed above, Fig.~\ref{fig:width}a
shows this as a function of $t_{\rm{samp}}$ for the model above (with $q=11/6$,
$m_{\rm{max}}=3.16 \times 10^{22}$~g and $\dot{M}_{\rm{in}}=10^{10}$~g\,s$^{-1}$),
as well as for the same model but for accretion from mass distributions with different slopes 
$q$.
Clearly, how the median accretion rate varies with sampling time is a strong
function of that slope.
To understand why, we apply a simple model in which the median accretion rate is
approximated as that continuously accreted from objects smaller than $m_{\rm{tr}}$,
which would result in
\begin{equation}
  \dot{M}_{\rm{med}} = (m_{\rm{max}}/t_{\rm{samp}})n_{\rm{max}}^{1/(q-1)}.
  \label{eq:mmed}
\end{equation}
This equation would hold for $t_{\rm{samp}}<t_{\rm{samp,crit}}$, but for longer
sampling times, the median accretion rate would be the mean accretion rate $\dot{M}_{\rm{in}}$.
Despite its simplicity, Fig.~\ref{fig:width}a shows that this prescription
fits the Monte Carlo model reasonably well.
Thus we consider that the effect of stochasticity on such accretion measurements
is also well understood in the regime of accretion from a mass distribution, and that
we can extrapolate the results presented here to arbitrary sampling times, mean accretion
rates, maximum planetesimal masses, and power law indices, as indicated in
the top and right axes of Fig.~\ref{fig:width}a.

While Fig.~\ref{fig:width}a shows the median of the distribution, it
does not describe its width, which is characterised in Fig.~\ref{fig:width}b
using the range of accretion rates that cover the 90, 10 and 1\% points in the
distribution.
As was already evident from Fig.~\ref{fig:model3.5}b, as
long as $t_{\rm{samp}} \ll t_{\rm{samp,crit}}$ (i.e., $n_{\rm{max}} \ll 1$),
the width of this distribution is relatively constant and independent of
$t_{\rm{samp}}$. 
However, the breadth of the distribution also depends on $q$, with steeper mass distribution
slopes (larger $q$) resulting in narrower accretion rate distributions.

The predictions of the model for stochastic accretion from a mono-mass planetesimal
distribution are also plotted on Fig.~\ref{fig:width} (reproduced from
Fig.~\ref{fig:analytmono} with appropriate scaling).
This comparison shows that the incorporation of a distribution of masses for the
accreted material substantially changes the character of the accretion rate distribution that
would be measured.
One difference is that the median accretion rate changes much more slowly with sampling
time.
This is because, for short sampling times, there is not only mass present in the
atmosphere shortly after an accretion event, rather there is always mass in the
atmosphere, albeit at a slightly lower level, from the accretion of small objects
in the distribution.
Another difference is that the accretion rate distribution is much narrower, because there
is always a plentiful supply of small objects to maintain the mass in the atmosphere at
a steady level, even if larger objects can still be accreted leading to increased
mass levels.

\subsection{Population model}
\label{ss:representative}
Having characterised what the distribution looks like for a single
accretion rate, it is relatively simple to determine what kind of
population model would be needed to fit the data.

\subsubsection{Constraint on $m_{\rm{max}}$, $\mu$ and $\sigma$}
\label{sss:mmax}
First of all we can use the arguments of \S \ref{ss:monomono} to 
rule out the mono-rate model by looking at the long sinking time bin.
This is because Fig.~\ref{fig:width}b shows that the factor of $\sim 1000$
between the inferred accretion rate at the 10\% and 50\% points in the distribution
cannot be achieved without a mass distribution with a very small value of $q$.
This would be equivalent to having a mono-mass distribution, which was ruled out
from the shorter sinking time bins in \S \ref{ss:monomono}.
Thus, as in \S \ref{ss:monomulti}, we assume a log-normal distribution for
$\dot{M}_{\rm{in}}$ parameterised by $\mu$ and $\sigma$.

The distribution of accretion rates in the long sinking time bin is not
indicative of the shape of the mass distribution, rather it is more likely representative
of the distribution of input accretion rates (as surmised in \S \ref{ss:monomulti}).
The correspondence is not necessarily exact, as the input accretion rates could be
higher than this.
Indeed, if the maximum planetesimal mass was large enough so that
$n_{\rm{max}}<1$ for $t_{\rm{sink}} \approx 1$~Myr and
$\dot{M}_{\rm{in}} \approx 10^{10}$~g\,s$^{-1}$, then
the accretion rates would on average be inferred to be lower than the
input rates for all stars in this bin (and for all stars in all bins);
this corresponds to a maximum planetesimal mass of $>3.2 \times 10^{23}$~g.
While \S \ref{ss:monomulti} used this argument to set a upper limit on $m_{\rm{p}}$
that is even lower than this, such a constraint is not necessary here.
This is because, although the distribution of inferred rates would be broader than that of
the input rates when $n_{\rm{max}}<1$, the mass distribution limits the effect of
broadening to a level that depends on $q$ (Fig.~\ref{fig:width}b), and it is only
extremely small values of $q$ (i.e., mono-mass distributions) for which that broadening
is so great that it is required that $n_{\rm{max}}>1$ to curtail it.

In fact, it turns out to be necessary in this instance for the maximum planetesimal
mass to be larger than the limit given in the last paragraph.
If it were much lower than this, then it would still be possible to construct an input
accretion rate distribution that allows a reasonable fit to the long sinking time bin
(e.g., for small enough $m_{\rm{max}}$ this would be
$\mu=6.6$ and $\sigma=1.5$ as discussed in \S \ref{ss:monomulti}).
However, a significant fraction of the white dwarfs in the shorter sinking time bins
would have $n_{\rm{max}}>1$ and so would have inferred accretion rates that are
indistinguishable from those in the long sinking time bin.
This issue is just becoming evident in Fig.~\ref{fig:monomulti}c, where
the model distributions for the long and medium sinking time bins are
indistinguishable for $\dot{M}>10^9$~g\,s$^{-1}$.

On the other hand, as long as $m_{\rm{max}}$ is above this limit, then its value does not
affect the quality of the fit.
This is because, as long as $n_{\rm{max}} \ll 1$ for all white dwarfs, the distribution
of accretion rates measured on each is the same if $\dot{M}_{\rm{in}}m_{\rm{max}}^{q-2}$
is kept constant (see eq.~\ref{eq:mmed} and Fig.~\ref{fig:width}).
That is, if we increase $m_{\rm{max}}$ above this limit, we can ensure that the model retains the
same accretion rate distributions for the different sinking time bins by also increasing
$\mu$ proportionately.
Thus, we will set $m_{\rm{max}}=3.2 \times 10^{24}$~g, i.e., a factor of 10 above this
limit, which is comparable with the mass of the largest Kuiper belt objects, noting that
lower values may be possible, as long as they are accompanied by lower input accretion rates,
though we expect the fit to the shorter sinking time bins to deteriorate as $m_{\rm{max}}$ is
decreased.

Note that another consequence of all white dwarfs accreting with $n_{\rm{max}}<1$ is that the
distributions of observationally inferred accretion rates should have very similar shapes
for the different sinking time bins.
It is just their median levels that would be offset by an amount that can be estimated from
eq.~(\ref{eq:mmed})
\begin{equation} 
  \dot{M}_{\rm{med}} \propto t_{\rm{samp,med}}^{(2-q)/(q-1)},
  \label{eq:mmed2}
\end{equation}
where $t_{\rm{samp,med}}$ is the median sampling time in the bin.
This is true as long as this width is not dictated by the width of sampling times within the bin,
since this means that the bins have the same width in their distribution of $n_{\rm{max}}$.

Practically we proceed with the modelling by assuming a value for $q$ (and for $m_{\rm{max}}$),
and then constraining the parameters of the input accretion rate distribution $\mu$ and $\sigma$
from a fit to the long sinking time bin.

\subsubsection{Constraint on $q$ and $t_{\rm{disc}}$}
\label{sss:q}
Given that the input accretion rate distribution can be chosen to provide
a reasonable fit to the long sinking time bin, the shorter sinking time bins can be used
consecutively to determine the parameters $q$ and $t_{\rm{disc}}$.
For example, the medium sinking time bin has the 15\% point in its distribution
a factor of $R_{\dot{M}} \approx 0.036$ lower in accretion rate than that of the
long bin.
Since the median sinking times of these bins are 850~yr and 0.37~Myr and so
have a ratio $R_{\rm{t}}=0.0024$, then eq.~(\ref{eq:mmed2}) shows that, ignoring
any effect of disc lifetime, this would require
\begin{equation}
  q = (2+X_{\rm{R}})/(1+X_{\rm{R}}), \label{eq:q}
\end{equation}
where $X_{\rm{R}} = \log{R_{\dot{M}}}/\log{R_{\rm{t}}} \approx 0.56$.
That is, to get a simultaneous fit to these bins would imply $q \approx 1.64$.

The problem is that the same argument cannot apply to the short sinking time bin, since the
median sinking time in this bin is 0.017~yr, which is $R_{\rm{t}}=4.7 \times 10^{-8}$ lower
than that of the long bin, which means that its accretion rate distribution should
be $\sim 10^{-5}$ times lower than that of the long bin.
Although the paucity of detections in the short sinking time bin (due to the poorer detection
threshold for these white dwarfs that are necessarily younger and hotter) means that 
its distribution is not well known, the fact that there are any detections
at all seems to rule this out.
However, the distribution inferred from the observations is readily accounted for if the
accretion is mediated through a disc as discussed in \S \ref{ss:tdisc},
since this would increase the effective sampling time (eq.~\ref{eq:tsamp}),
exclusively in the short bin for a suitably chosen disc lifetime.
Given that the inferred accretion rate distribution is poorly defined observationally, any
estimate of the disc lifetime on this basis would have significant uncertainty.
To make progress we note that the different sinking time bins should have distributions
that are offset in accretion rate by a factor of around $t_{\rm{samp}}^{X_{\rm{R}}}$
(see eq.~\ref{eq:mmed2}).
Thus to get the 5\% points in the short and long sinking time
bins offset by $\sim 10^{-3}$ would require the short bin to have a median sampling
time of around 2~yr.

\begin{figure}
  \begin{center}
    \vspace{-0.1in}
    \begin{tabular}{c}
      \hspace{-0.2in}
      \psfig{figure=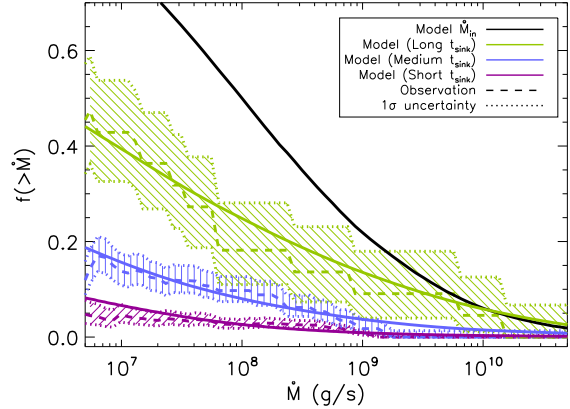,height=2.4in}
    \end{tabular}
    \caption{Population model fit to the observationally inferred accretion
    rate distributions for a model in which
    the accreted material has a mass distribution defined by
    $m_{\rm{max}}=3.2 \times 10^{24}$~g and $q=1.57$, the input accretion rates
    have a log-normal distribution defined by $\mu=8.0$ and $\sigma=1.3$,
    and a disc lifetime of $t_{\rm{disc}}=20$~yr was also assumed.
    }
   \label{fig:pop}
  \end{center}
\end{figure}

Combining these previous estimates, and making small (consecutive) adjustments to improve the
fit, we show in Fig.~\ref{fig:pop} the predictions for a population model with the
following parameters: $m_{\rm{max}}=3.2 \times 10^{24}$~g,
$\mu=8.0$, $\sigma=1.3$, $q=1.57$, $t_{\rm{disc}}=20$~yr.
Given the arguments in the preceding paragraphs it is not surprising that this
provides a reasonable qualitative fit to the observationally inferred accretion
rate distributions, including how those distributions
differ between the sinking time bins.
Quantitatively the fit is also good, with $\chi_s^2=0.8,2.9,1.7$ for the long, medium
and short sinking time bins, respectively.
We prefer not to give formal uncertainties on the model parameters, since this gives
the impression that they are better constrained than they really are,
and ignores the (still quite significant) uncertainty in the rate distributions
inferred from the observations,
as well as the systematic uncertainty on whether the
model includes all of the relevant physics.
Rather the intention here is to show how the model behaves, to show that it
provides a qualitatively reasonable fit to the observationally inferred
accretion rate distributions, and to motivate further
observations that provide better constraints on these distributions, and their
dependence on sinking time and cooling age.
Nevertheless, the discussion above illustrates the various degeneracies between the
different parameters, and also gives a feeling for how changing these parameters would
affect the quality of the fit.

\section{Discussion}
\label{s:disc}
Thus far we have discussed how the stochastic accretion of a mono-mass planetesimal
distribution would be manifested in observations of white dwarfs with different
sinking times (\S \ref{s:mod}), as well as the effect of allowing the mass distribution
to encompass a range of masses (\S \ref{ss:params}).
These models were then compared with observations of metal pollution onto white dwarfs
that were used to get inferred accretion rate distributions as summarised in \S \ref{s:obs}.
It was shown that, although the mono-mass distribution can provide a reasonable fit to these
distributions, it does not reproduce the correct qualitative differences with
sinking time (\S \ref{s:mono}), while allowing a mass distribution results in a
much better fit and the correct qualitative behaviour (\S \ref{ss:representative}).
This section considers the implications of the model results.
To start with in \S \ref{ss:phys} we outline a physical picture for the evolution
of the material that is accreted by the white dwarfs.
In \S \ref{ss:implications} we consider how the parameters derived in the previous
sections fit in with that picture.
We also consider caveats to the model, and how it might need to be improved upon in
the future (\S \ref{ss:caveats}), as well as observational avenues to further
constrain what is going on (\S \ref{ss:newobs}).

\subsection{Physical model of accretion onto white dwarfs}
\label{ss:phys}
Consider that the main sequence progenitor of the white dwarf had an
orbiting belt of planetesimals, like the Solar System's asteroid or Kuiper belts.
If their orbits were stable for at least several hundred Myr this belt would
have survived to the post main sequence (e.g., Greenstein 1974).
Unless the belt is very low mass, and assuming the collision velocity is
high enough (Heng \& Tremaine 2010), it is inevitable that mutual collisions
amongst the planetesimals would have set up a collisional cascade.
While this would have reduced the belt mass, it also means that its
mass distribution would be reasonably well-defined.
For example, for the ideal case where the planetesimals' dispersal threshold
is independent of mass it would be expected that $q=11/6$.
For the more realistic case that the dispersal threshold is mass dependent,
this would result in a distribution with a slightly different index, or one
in which the distribution exhibits different indices in different mass ranges
(e.g., O'Brien \& Greenberg 2003; Wyatt, Clarke \& Booth 2011).

When the star evolved to become a white dwarf some mechanism could perturb the orbits so
that material is scattered from the belt toward the star.
Like the comets in the Solar System, some of this material would end up being
accreted onto planets or ejected into interstellar space or scattered into an Oort Cloud
analogue.
However some fraction could end up incorporated into a disc that accretes onto the star.
One candidate for the perturbing process is that the inner edge of the belt
was located at the edge of the chaotic region of resonance overlap of an interior
planet, and that stellar mass loss caused the size of that unstable region to
expand (Bonsor, Mustill \& Wyatt 2011).
Another possibility is that a planet lies exterior to the belt, and that one of
its resonances lies in the middle of the belt;
the resonance is unstable, and so empty on the main sequence (similar to the
Kirkwood gaps in the asteroid belt), but stellar mass loss causes the resonance
to expand feeding mass into the dynamically unstable region (Debes et al. 2012). 

Both of the mechanisms discussed above are dynamical and affect all material in the belt
regardless of mass.
This means that the mass that is scattered would retain the mass distribution of the belt,
and it might be expected that the mass distribution of accreted material
(i.e., that described in eq.~\ref{eq:nd}) is indicative of the mass distribution of the belt.
However, there are physical processes that might bias the scattering
process to different masses;
e.g. sub-km planetesimals could have been dragged in by stellar wind drag during the
AGB phase (Bonsor \& Wyatt 2010), perhaps also getting trapped in planetary
resonances as a result (Dong et al. 2010).
The efficiency of tidal disruption, or of subsequent incorporation into a disc,
may also have some dependency on the mass of the original object, so that the
mass distribution of accreted material may differ from that of the belt.

\subsection{Implications of model fits}
\label{ss:implications}
Given the physical picture of \S \ref{ss:phys} we can now discuss whether the model
parameters required to fit the observationally inferred accretion rate distributions 
(i.e., those given in \S \ref{ss:representative}) are physically plausible.

\textit{Accretion rate distribution ($\mu$ and $\sigma$):}
The inferred accretion rate distribution is remarkably similar to that
derived in the model of Bonsor et al. (2011).
That model used the well characterised population of main sequence A star
debris discs (Wyatt et al. 2007), determined what that population would look
like at the start of the white dwarf phase (Bonsor \& Wyatt 2010), then
considered the fraction of mass scattered due to the increase of resonance
overlap due to stellar mass loss if the discs were truncated at the inner edge by a
planet.
A fixed fraction of the scattered mass ($0.6$\% based on simulations of the Solar System)
was assumed to make it onto the star, and Figs 7-8 of their paper predict the distribution of
mass accretion rates experienced by white dwarfs as a function of age.
Here we fit the Bonsor et al. (2011) results with log-normal distributions to
get the appropriate $\mu$ and $\sigma$ at different ages.
We find that white dwarfs in the age range 10-5000~Myr in their model have an approximately
log-normal distribution of accretion rates with $\mu=8.1$ and $\sigma=1.6$, coincidentally
almost identical to that in our model. 
There is a slight dependence on age in their model, in that the median decreases
$\propto t_{\rm{age}}^{-1.1}$, though we show in \S \ref{ss:caveats} that this is still
consistent with the observationally inferred accretion rates in \S \ref{ss:distrn}.
While the agreement with our results should not be taken as strong support for
the Bonsor et al. (2011) model --- indeed the other models for the origin of the accretion
(e.g., Debes et al. 2012) may reproduce a similar distribution of accretion rates ---
it does at least mean that the required rates are at a level, and have a width in
their distribution, that is physically plausible.

\textit{Mass distribution ($m_{\rm{max}}$ and $q$):} 
The mass distribution required to fit the observationally inferred
accretion rate distributions is in-line with
the distribution expected for the parent belt due to collisional evolution.
Although $q=11/6$ is that expected in an infinite cascade of planetesimals
with dispersal threshold that is independent on mass (Dohnanyi 1969),
it is expected that planetesimals more massive than $10^{12}$~g have strengths that increase
with mass, i.e. $Q_{\rm{D}}^\star \propto m^{b}$, due
to self gravity so that $q=(11+3b)/(6+3b)$ (O'Brien \& Greenberg 2003; Wyatt et al. 2011). 
Typically $b \approx 1/2$ in this regime giving $q \approx 5/3$
(e.g., Benz \& Asphaug 1999; L\"{o}hne et al. 2008).
This is a reasonable approximation for the asteroid belt, which is thought
to have reached collisional equilibrium, at least for objects smaller than
around $3 \times 10^{21}$~g,
before it was depleted to its currently low level
(e.g., Durda, Greenberg \& Jedicke 1998; Bottke et al. 2005).
However, the mass distribution in the Kuiper belt is weighted more to smaller
objects (see Vitense et al. 2010),
likely because it retains the primordial distribution;
this is closer to the distribution with $q=2$ assumed in the models
of Jura (2008).
Thus one implication of the low value of $q=1.57$ inferred here could
be that the accretion onto white dwarfs originates in a collisionally evolved population.
This is perhaps unsurprising, since this only rules out planetesimal belts
that are very far from the star (for which collisional evolution timescales are long), or
those that were depleted in dynamical instabilities, both of which might be expected
to be unfavourable to high accretion rates.
However, it is not clear that $q$ in this model corresponds exactly
with that in the parent belt, since some mass ranges could be more readily
incorporated into an accretion disc.

Having constrained the mass distribution it is also helpful to remind the reader that 
in the model most stars are accreting continuously from objects in the mass
distribution up to a mass $m_{\rm{tr}}$.
Although that mass varies from star to star, it is possible to obtain a feeling for
what objects are contributing to the observations by considering that the model's
behaviour can be well explained by assuming that the accretion rate that is inferred
for a given star is its median level, which is $m_{\rm{tr}}/t_{\rm{samp}}$ (see eq.~\ref{eq:mmed}).
This means that, perhaps unsurprisingly, $m_{\rm{tr}} = \dot{M}_{\rm{obs}}t_{\rm{samp}}$ and
so is usually roughly the mass in pollutants in the white dwarf's convection zone
(except for the DAs with sinking times that are shorter than the disc lifetime).
So, for commonly inferred accretion rates of around $10^8$~g\,s$^{-1}$, such accretion
levels for a typical star in the long sinking time bin (i.e., one with the median sampling
time of 0.37~Myr), would be dominated by the accretion of
$1.2 \times 10^{21}$~g objects,
those in the medium bin (with a median sampling time of 850~yr) by $2.7 \times 10^{18}$~g objects,
and those in the short bin (with a median sampling time of 20~yr set by the disc lifetime)
by $6.3 \times 10^{16}$~g objects;
these values correspond to 91, 12 and 3.4~km diameter objects, respectively, for a density of 3~g\,cm$^{-3}$.
However, note that since inferred accretion rates span roughly 4 orders of magnitude, with a similar
range in sampling times, these values should only be considered representative for each bin.
Considering the polluted DAs in our sample we find that $m_{\rm{tr}}$ covers the range
$2.3 \times 10^{15}$ to $7.1 \times 10^{19}$~g (with a median of $5.2 \times 10^{17}$~g),
while that for polluted non-DAs covers the range
$6.2 \times 10^{19}$ to $1.3 \times 10^{23}$~g (with a median of $9.6 \times 10^{20}$~g).
This means that there is a continuous range of $m_{\rm{tr}}$ spanning 8 orders of magnitude in
mass, but that the planetesimals dominating the pollution on DAs and non-DAs are smaller and larger
respectively than $\sim 6.6 \times 10^{19}$~g (i.e., $\sim 35$~km diameter for 3~g\,cm$^{-3}$).
To understand why there is (coincidentally) such a neat division between DAs and non-DAs,
it is helpful to note that lines of constant $m_{\rm{tr}}$ are horizontal on
Fig.~\ref{fig:obs}c up to $t_{\rm{sink}}=t_{\rm{disc}}$ (i.e., 20~yr in our best fit model),
then fall off $\propto t_{\rm{sink}}^{-1}$ for $t_{\rm{sink}}>t_{\rm{disc}}$.

Similar logic can be used to estimate the input accretion rate as
\begin{equation}
  \dot{M}_{\rm{in}}/\dot{M}_{\rm{obs}} = (\dot{M}_{\rm{obs}} t_{\rm{samp}}/m_{\rm{max}})^{q-2}, 
  \label{eq:mdotinobs}
\end{equation}
where the quantity in brackets is the number of objects of mass $m_{\rm{max}}$ that would have to
be accreted per sampling time to reproduce the inferred rate.
To give a couple of specific examples:
\begin{itemize}
\item The $\sim 10^{21}$~g of metals in the atmosphere
of the prototypically polluted non-DA white dwarf vMa2, that has a sinking time of $\sim 3$~Myr,
gives an inferred accretion rate of $\sim 10^7$~g\,s$^{-1}$.
In this model, equation~(\ref{eq:mdotinobs}) says that the input accretion rate is likely
to be around 30 times higher than that inferred from the observations, which would put it
slightly above the median input rate in the model;
the pollutants currently in the atmosphere arrived in multiple accretion events
of planetesimals smaller than $\sim 84$~km.
\item The $10^{16}$~g of metals in the atmosphere of the prototypical DA white dwarf
G29-38, that has a sinking time of $<1$~yr,
gives a higher than average inferred accretion rate of $\sim 10^9$~g\,s$^{-1}$.
Equation~(\ref{eq:mdotinobs}) implies that this star has a much higher than average input accretion
rate of $\sim 6 \times 10^{11}$~g\,s$^{-1}$, which would put it in the top 0.2\% of
accretion rates,
and its pollutants arrived in multiple accretion events of $\leq 6$~km planetesimals.
Note that accretion levels above $10^9$~g\,s$^{-1}$ in the short sinking time bin
occur around just $0.8^{+0.6}_{-0.4}$\% of the sample, both in the distributions inferred
observationally and in the model,
consistent with the above consideration of the rarity of this object. 
\end{itemize}

\textit{Disc lifetime:}
As discussed in \S \ref{ss:representative}, the constraints on the disc lifetime are
not very stringent. 
We know that it must be shorter than $\sim 0.1$~Myr for there to be a dependence on
sinking time (since otherwise all stars would have the same sampling time irrespective
of their sinking time), and further that it must be shorter than around 1000~yr if we want there
to be a difference between the two shorter sinking time bins.
A disc lifetime as short as 20~yr, if confirmed in later analysis, would have significant
implications for the physics of the disc accretion, but we do not discuss this further here
given the simple way in which disc lifetime was included in this model, and that any 
statement on the disc lifetime should also take into account observational evidence concerning
the disc itself (e.g., regarding near-IR excesses or gas).

\begin{figure}
  \begin{center}
    \vspace{-0.1in}
    \begin{tabular}{c}
      \hspace{-0.2in}
      \psfig{figure=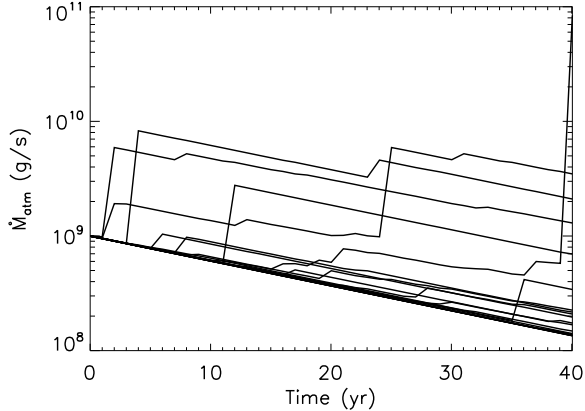,height=2.4in}
    \end{tabular}
    \caption{Evolution of the inferred accretion rate for 20 stars from the model of
    Fig.~\ref{fig:pop} that fit the characteristics of G29-38
    ($t_{\rm{sink}} \approx 1$~yr and $\dot{M}_{\rm{atm}}=10^9$~g\,s$^{-1}$
    at the present day).
    }
   \label{fig:g2938}
  \end{center}
\end{figure}

It should be noted that a disc lifetime of 20~yr does not necessarily mean that
pollution levels are expected to decay on such timescales.
While the signature of individual events would decay on the sampling timescale
(eq.~\ref{eq:tsamp}), the pollution levels of many stars in this model are maintained 
indefinitely from multiple accretion events (although individual events can result
in temporarily higher levels).
To quantify this, Fig.~\ref{fig:g2938} shows the predicted evolution of the accretion
signatures for twenty stars from the best fit model population of Fig.~\ref{fig:pop}
that fit the characteristics of G29-38, that is currently 
inferred to be accreting at a rate of $10^9$~g\,s$^{-1}$ and has a sinking time of $<1$~yr.
This illustrates how for many stars like G29-38 the accretion is
maintained at a similar level over decadal timescales (consistent with
the decades-long persistence of circumstellar dust and pollution toward
G29-38: Zuckerman \& Becklin 1987; Koester et al. 1997; Hoard et al. 2013),
but that a rare few will undergo further brightening events, while a larger fraction
will decay monotonically over the disc timescale.
A similar plot for vMa2, for which pollution in its atmosphere has persisted for around a
century (van Maanen 1917; Dufour et al. 2007), predicts no discernable evolution on such timescales
due to the $\sim 3$~Myr sinking time for this non-DA star.

Since the disc timescale is not well constrained in this model, we note that monitoring
variability in accretion signatures on DA white dwarfs would be an excellent way
to set constraints on this parameter.
Variability in the accretion has been suggested for G29-38 based on a 70\%
increase in atmospheric pollution over a two year timescale (von Hippel \& Thompson 2007),
though other studies found the pollution level to have remained constant over
timescales of days to years (Debes \& L\'{o}pez-Morales 2008), and mid-IR
observations of the circumstellar disk have also been shown to remain
constant over yearly timescales (Reach et al. 2009).
Since $18/20$ models on Fig.~\ref{fig:g2938} remain within a factor
of 2 of $10^9$~g\,s$^{-1}$ over the first decade, but $3/20$ undergo
moderate levels of brightening (10-200\% increases) in this period,
we conclude that the model is consistent with the observations.

\subsection{Caveats}
\label{ss:caveats}
Several assumptions are implicit in the model, and were included not necessarily
because these are the most physically realistic, rather to minimise model parameters
that would be impossible to constrain.

We assumed that the accretion rate is constant and independent of age, justified by the
lack of evidence for any age dependence in the observationally inferred accretion rates.
However, in \S \ref{ss:implications} we noted that one of the physical models,
and perhaps all of them (e.g., Veras et al. 2013), require some fall-off with age.
It would be relatively simple to include an age dependence in the model;
for example the model population could have both the same sinking time distribution
and the same cooling age distribution as the observed population.
One consequence of this would likely be that we would infer a steeper slope (higher $q$) for
the mass distribution.
This is because of the point noted at the end of \S \ref{ss:distrn} that a decrease in
accretion rate with age would mean that the accretion rate distribution in the medium
sinking time bin would have been higher relative to the other two bins had it been measured
at a comparable age.

It should also be possible to use the lack of age dependence in the observationally
inferred accretion rates to set
constraints on the evolution of the input accretion rate.
For example, the Bonsor et al. (2011) model results in a median input accretion rate
that decreases by a factor of $0.16$ between cooling age bins of 100-500~Myr
and 500-5000~Myr.
Equation~(\ref{eq:mmed}) shows that, for the inferred mass distribution index $q$,
the consequence for the distribution of accretion rates would be to shift those in
the older bin to lower values by a factor of $0.16^{1/(q-1)} \approx 0.04$.
However, the same equation shows that this difference is counteracted by
the 37 times higher median sampling time in the older cooling age bin
(730~yr compared with 20~yr in the younger bin), which would be expected to increase
this by $37^{(2-q)/(q-1)} \approx 15$.
In other words the net result would be for the older bin to appear to have
accretion rates that are very similar to those in the younger bin, in
agreement with Fig.~\ref{fig:obs}b.
The low significance trend in Fig.~\ref{fig:obs}b could argue against an
evolution in accretion rates that is significantly faster than that of
the Bonsor et al. (2011) model, but we consider that interpretation of any
evolutionary signal is complicated by the sinking time dependence and so no
strong statements can be made at this stage.

The mass distribution was assumed to extend up to objects nearly as massive as Pluto
for all stars, and to have a single slope across all masses.
The large value of $m_{\rm{max}}$ is not necessarily a problem, given the prevalence
of debris discs.
However, it should be noted that there is no requirement for objects larger
than a few km in most debris discs (e.g., Wyatt \& Dent 2002), though the presence of
objects the mass of Pluto would provide a natural explanation for the origin of
the disc stirring (Kenyon \& Bromley 2004).
Nevertheless there remains a discontinuity between the parameters derived here and
the model of Bonsor et al. (2011), which was itself based on a model in which the
maximum planetesimal size was nominally 2~km and
the distribution had a slope of $q=11/6$ (Bonsor \& Wyatt 2010).
However, it should be noted that the
Bonsor \& Wyatt (2010) model parameters are only meant to be representative values, given
its simplistic prescription for collisional evolution, and that
more realistic models for collisional evolution on the main sequence
would be expected to provide similar results for larger maximum planetesimal masses
and with shallower slopes in the relevant regime (e.g., L\"{o}hne
et al. 2008).
Certainly future models could readily include a mass dependent planetesimal strength
(e.g., Wyatt et al. 2011), and the results incorporated into the
Bonsor et al. (2010) model
to make a revised prediction, though we would not anticipate the conclusions
of this paper to be affected in any substantial way.

The implicit assumption that disc lifetime is independent of
the accretion rate, of the mass of objects being accreted, and of disc mass, is more
of a concern.
If discs last longer when the accretion rate is larger, for example because it
takes longer to break down large objects into dust, this would bias the
observations toward detecting the most massive events (and vice versa).
This is not a topic that we can cover adequately here, so we leave it for a future
paper.

Another factor of concern noted in \S \ref{ss:thermohaline} is the uncertainty
in the observationally inferred accretion rates, and in particular whether unmodelled
processes that act on the material after it has been accreted, such as thermohaline
convection in the stellar interior, would change the inferred accretion rates enough
to come to different conclusions on how their distribution depends on sinking time.
Assuming it is possible to approximate the signature of an accreted planetesimal in
a white dwarf atmosphere with exponential decay (eq.~\ref{eq:matmi}), albeit with a sinking time
modified from that of gravitational settling, then the model presented here will still
apply, and the implications of revised inferred accretion rate distributions can
be understood within the context of the arguments in this paper.
For example, as mentioned in \S \ref{sss:monomultitdisc}, if the inferred accretion rate
distributions turn out to be independent of (effective) sinking time, this would argue for a
disc lifetime $\gg 0.1$~Myr, effectively negating the importance of sinking time on how
accretion is measured.
This would mean that information on the mass distribution of accreted material
cannot be gleaned from the inferred accretion rate distributions.

\subsection{Future observations}
\label{ss:newobs}
One of the conclusions to arise from this analysis is to emphasise to observers that
non-detections have equal value to detections in our interpretation of this
phenomenon.
That is, we urge that future observations are presented as the distribution of inferred
accretion rates, and that these observations seek to quantify how those distributions vary with
sinking time, and to search for evidence of a dependence on cooling age.
It is premature to claim that the model can make testable predictions.
Although Fig.~\ref{fig:pop} does make a prediction for how the accretion rate
distributions extend to lower accretion levels, it should be recognised that the model
has sufficient free parameters to fit other distributions should they arise from the
observations, so a poor fit cannot be used to rule the model out.
Rather, what we do claim is that a dependence of the inferred accretion rate distribution
on sinking time is a natural consequence of the stochastic nature of accretion processes,
and that we can learn about the mass distribution of accreted material, and also about the disc
lifetime, by constraining those distributions through observations.
For example, confirming that the accretion rate distributions exhibit progressively lower
rates for shorter sinking times, with differences that are much smaller than the many orders
of magnitude difference in sinking times, would strengthen the conclusion that the accretion
arises from planetesimals with a wide range of masses, rather than from a mono-mass
planetesimal distribution.
Observations of the variability of accretion signatures on individual DA white dwarfs
can also be informative of the disc lifetime (see Fig.~\ref{fig:g2938}).

Another promising avenue for comparison of the model with observations is to consider
its compatibility with the fraction of stars with detectable accretion that have infrared excess.
For example, 10/21 DA white dwarfs with accretion rates inferred at $>10^8$~g\,s$^{-1}$ have
infrared excess, whereas this fraction is lower at 7/30 for DB white dwarfs (Girven et al. 2012).
This fits qualitatively within the context of the model presented here.
Given the short lifetime of the disc, its luminosity would be expected to be set by the accretion
of planetesimals in a similar mass range to those that dominate the atmospheric pollution of stars
in the short sinking time bin.
It is a relatively small fraction of the DAs that have accretion rates inferred to be
$>10^8$~g\,s$^{-1}$, and those that do are likely to be those that have atypically large input
accretion rates, and so it might be expected that these also have bright discs.
However, it is a relatively large fraction of the DBs that have accretion rates inferred at
this level, and so their input accretion rates would be expected to span a lower range
than the DAs detected at this level, which would explain why their discs are fainter on average.
Although this is qualitatively reasonable, such comparisons should be made
more quantitatively along with a more detailed consideration of the
observations, and a more detailed prescription for the disc in the model.

It might appear that one way of testing this model would be to use the observationally
inferred accretion rates of different metals in the same star, since different metals have different sinking
timescales;
e.g., this model would predict that metals with longer sinking timescales
originate in the accretion of (on average) more massive objects and so should exhibit a higher
inferred accretion rate.
However, Fig.~\ref{fig:obs}a shows that sinking times vary only by a factor of a few for
different metals.
Furthermore, this analysis would only be appropriate if the abundance of the material being
accreted was known, since otherwise differences could be explained by compositional
variations.
Thus observations of different metals in the same star are usually used to determine
the abundance of the accreted material, rather than to make inferences about the
accretion process.
However, Montgomery, Thompson \& von Hippel (2008) show how measuring different
variability patterns in the accretion rates of different metals could be used to
set constraints on gravitational settling times.

\section{Conclusion}
\label{s:conc}
This paper explores the effect of stochastic processes on
measurements of accretion of planetesimals onto white dwarfs.
We first quantified the distribution of accretion rates inferred from observations of atmospheric
pollution in \S \ref{s:obs}.
As previous authors had found, we concluded that this distribution has a
dependence on the timescale for metals to sink in the atmosphere, with tentative evidence
that our sample could be split into three sinking time bins with accretion rates that are
progressively lower as sinking time is reduced;
there was no evidence for a dependence on cooling age.
These conclusions use the typical assumption that gravitational settling is
the dominant process removing metals from the atmosphere.
As such they should be revisited once the effect of thermohaline convection has been fully
characterised.

In \S \ref{s:mod} we showed how the accretion of a mono-mass population of planetesimals
would be manifested in observations of atmospheric pollution.
We described the resulting distribution of inferred accretion rates both analytically and using
a Monte Carlo model, demonstrating how stochastic processes cause that distribution to
have a strong dependence on sinking time.
We compared this model to the observationally inferred accretion rate distributions
in \S \ref{s:mono} to find
that while it is easy to reproduce the distributions inferred for one (or with more effort two)
of the sinking time bins, a concurrent fit to the distributions in all three sinking time bins
is not possible.
The problem is that the many orders of magnitude difference in sinking time between
white dwarfs would cause this model to have larger differences in inferred accretion rates
than determined from the observations.
The model of this ilk that most closely matches the observationally inferred distributions
invoked a disc lifetime of 0.01-0.1~Myr to smooth out the accretion
that is measured on stars with short sinking times.

In \S \ref{s:mod2} we showed how allowing the accreted planetesimals to have a range of masses
substantially changes the accretion rate distribution that would be inferred, with a
less dramatic dependence of inferred accretion rates on sinking time more in-line with the
observationally inferred accretion rates.
There is also an important conceptual difference.
With a mono-mass planetesimal distribution, stars only
exhibit accretion signatures shortly after a planetesimal has been accreted. 
However, when there is a distribution of planetesimal masses, stars always exhibit
accretion signatures, because small enough planetesimals (quantified in eq.~\ref{eq:dtr})
are being accreted continuously.
We show that such a model provides a good fit to the
observationally inferred distributions with a relatively shallow mass distribution for the
accreted material ($q=1.57$), similar to that expected for a
collisionally evolved population.
Along with the other parameters of the model (e.g., the input accretion rate
distribution) we find that the atmospheric pollution signatures are consistent with
the accretion of the descendants of the debris discs seen around main sequence
stars, as predicted by Bonsor et al. (2011).
There are however several outstanding questions, such as the origin and nature
of the disc through which the accretion takes place.

If this interpretation is backed up with future observations,
including a consideration of the importance of thermohaline convection on
the inferred accretion rates, one implication is that atmospheric
pollution does not always originate in the stochastic accretion of individual objects;
i.e., pollution in non-DA white dwarfs is not necessarily a historical relic of past events
(Farihi et al. 2012b).
While the accretion of individual objects can affect the accretion signature, for many
stars this is dominated by the continuous accretion of moderately sized planetesimals.
The closest model in the literature to that presented here is the two population model
of Jura (2008), in which large objects are accreted infrequently and small objects are
accreted continuously.
However, the role of stochastic processes is more subtle than suggested by that model,
since the transition between these two populations occurs at different planetesimal
masses for different stars, and this is what imprints the mass distribution on the
distribution of observationally inferred accretion rates.

Finally we note that the insight gleaned from this study into the way stochastic
processes are manifested in observations may also relevant to other fields of
astrophysics.
For example, the origin of exozodiacal emission from the scattering of cometary
material into the inner regions of nearby planetary systems is a process
that has many analogies to that studied here (Bonsor, Augereau \& Thebault 2012).

\appendix
\section{Application of shot noise to accretion onto white dwarfs}
\label{a:shot}
We are interested in computing the distribution of the mass of material
still in the white dwarf atmosphere, $M_{\rm{atm}}$, as a result of the accretion
of a succession of planetesimals, each of mass $m_{\rm{p}}$, given that the 
mass in the atmosphere drains exponentially on a timescale $t_{\rm{sink}}$. 
This problem is analogous to various other problems in the current 
literature including dam theory (where dams are assumed filled by 
rainfall occurring stochastically), fluid queuing theory (which differs 
from ordinary queuing theory in that the number of customers in the queue 
can now be any real number and not just an integer), and risk theory in 
the computation of insurance claims.
The origins of these theories date back to the late 19th century where 
considerations of shot noise in electrical systems began.

What interests us here is the computation of the distribution of the amplitude of
shot noise, $I(t)$, caused by a superposition of impulses occurring at random 
Poisson distributed times $\cdots, t_{-1}, t_0, t_1, t_2, \cdots$, for 
the case in which all the impulses have the same shape, $F(t)$.
In this case
\begin{equation} 
  \label{sum}
  I(t) = \sum_i F(t-t_i),
\end{equation}
and for the problem of interest to us
\begin{equation}
  F(t) = H(t)e^{-t},
  \label{eq:appft}
\end{equation}
where $H(t)$ is the Heaviside step function.

Gilbert \& Pollak (1960) show that for the general case, the cumulative amplitude 
distribution function
\begin{equation}
  Q(I) = Pr[I(t) \le I]
\end{equation}
obeys an integral equation
\begin{equation}
  I\,Q(I) = \int^I_{-\infty} Q(x) \, dx  + n \int^\infty_{-\infty} Q[I-F(t)] \, F(t) \, dt,
  \label{eq:a4}
\end{equation}
where $n$ is the mean rate of arrival of shots. 

For exponential shots, given by eq.~(\ref{eq:appft}), this can be written as a 
difference differential equation for density $P(I) = dQ/dI$, {\em viz.}
\begin{equation}
  I \frac{dP}{dI} = (n-1) P(I) - n P(I-1),
  \label{eq:a5}
\end{equation}
with the convention that $P(I) = 0$ when $I < 0$.

They show that for $0 < I < 1$, 
\begin{equation}
  P(I) = \frac{e^{-n \gamma}}{\Gamma(n)} I^{n-1},
  \label{eq:a6}
\end{equation}
where $\gamma = 0.577215665 \ldots$ is Euler's constant,
and that for $I > 1$ the difference differential equation can be 
converted to an integral form
\begin{equation}
  P(I) = I^{n-1} \left[ \frac{e^{-n \gamma}}{\Gamma(n)} - n \int_1^I P(x-1) \, x^{-n} \, dx \right].
  \label{eq:a7}
\end{equation}

Although for general shot functions, $F(t)$, the computation of the 
distribution function can be problematic (see, for example, Lowen 1990
and Gubner 1996), the numerical computation of P(I) in this 
case is straightforward, as long as care is taken with the integrable 
singularity at $I=0$ for $0 < n < 1$, that is, when the mean interval 
between shot is greater than the exponential decay timescale.

For large values of $n$, $n \gg 1$, that is when a large number of shots 
occur during each decay timescale, one can make use of the asymptotic 
formulation known as Campbell's Theorem (Campbell 1909a, 1909b).
Campbell's Theorem states that for $n \gg 1$ the distribution 
asymptotically approaches that of a Gaussian or normal distribution with 
mean
\begin{equation}
  \overline{I} = n \int_{-\infty}^\infty F(t) \, dt,
  \label{eq:a8}
\end{equation}
and standard deviation, $\sigma$, given by
\begin{equation}
  \sigma^2 = n \int_{-\infty}^\infty [F(t)]^2 \, dt,
  \label{eq:a9}
\end{equation}
with error of order $1/n$.

Rice (1944) has generalised Cambpell's Theorem to the case when the shots have
a distribution of amplitudes, which corresponds in our problem to the case when
there is a distribution of asteroid masses.
In this case the amplitude $I(t)$ in equation~\ref{sum} becomes
\begin{equation}
  I(t) = \sum_i a_i F(t-t_i),
\end{equation}
where $\cdots a_1, a_2, a_3, \cdots$ are independent random variables 
all having the same distribution.

In this case for $n \gg 1$ the distribution approaches that of a 
Gaussian or normal distribution with mean
\begin{equation}
  \overline{I} = n  \, \overline{a} \int_{-\infty}^\infty F(t) \, dt,
\end{equation}
and standard deviation, $\sigma$, given by
\begin{equation}
  \sigma^2 = n \, \overline{ a^2} \int_{-\infty}^\infty [F(t)]^2 \, dt.
\end{equation}

\section*{Acknowledgments}
MCW acknowledges the support of the European Union through
ERC grant number 279973.
J. Farihi acknowledges support from the
STFC via an Ernest Rutherford Fellowship.


\end{document}